\documentclass[hidelinks,twocolumn,secnumarabic,amssymb,nobibnotes,floatfix,aps,prb,10pt]{revtex4-2}
\usepackage{amsmath}
\usepackage{amssymb}
\usepackage{gensymb}
\usepackage{esint}
\usepackage{physics}
\usepackage{graphicx}
\usepackage{multirow}
\usepackage{lipsum} 
\usepackage{hyperref}
\usepackage[T1]{fontenc}
\usepackage[utf8]{inputenc}
\usepackage{chemfig}
\usepackage{tikz}
\usetikzlibrary{calc}
\usetikzlibrary{backgrounds}
\usetikzlibrary{fit}
\usepackage{textcomp} 

\newcommand*\diff{\mathop{}\!\mathrm{d}}
\newcommand{\etal}{\textit{et al.}}
\newcommand{\LAMMPS}{\textsc{lammps}}
\newcommand{\drprobe}{\textsc{DrProbe}}
\newcommand{\fftw}{\textsc{fftw3}}
\newcommand{\fs}{\,\rm{fs}}
\newcommand{\ps}{\,\rm{ps}}

\usepackage{xcolor}
\hypersetup{
    colorlinks,
    linkcolor={blue!80!black},
    citecolor={blue!80!black},
    urlcolor={blue!80!black}
}
\definecolor{hlcolor}{RGB}{209, 21, 7}
\definecolor{movedcolor}{RGB}{22, 22, 200}

\newcommand{\highlight}[1]{#1}
\newcommand{\moved}[1]{#1}
\newcommand{\mhighlight}[1]{#1}
\newcommand{\fhighlight}[1]{#1}
\newcommand{\mmoved}[1]{#1}
\newcommand{\fmoved}[1]{#1}

\newlength{\minuslength}
\begin{document}
\title{Mode-Dependent Phonon Relaxation in fcc Ni: Insights from Molecular Dynamics Simulations with Frozen Trajectory Excitations}

\author{Wojciech Marciniak}%
    \email[email: ]{wojciech.marciniak@put.poznan.pl}%
    \affiliation{Department of Physics and Astronomy, Uppsala University, P.O. Box 516, 75120 Uppsala, Sweden}%
    \affiliation{Institute of Physics, Poznan University of Technology, Piotrowo 3, 60-965 Pozna\'n, Poland}%
    \affiliation{Institute of Molecular Physics, Polish Academy of Sciences,  M. Smoluchowskiego 17, 60-179 Pozna\'n, Poland}%
\author{Joanna Marciniak}%
    \affiliation{Department of Physics and Astronomy, Uppsala University, P.O. Box 516, 75120 Uppsala, Sweden}%
    \affiliation{Institute of Molecular Physics, Polish Academy of Sciences,  M. Smoluchowskiego 17, 60-179 Pozna\'n, Poland}%
\author{Jos\'e \'Angel Castellanos-Reyes}%
    \affiliation{Department of Physics and Astronomy, Uppsala University, P.O. Box 516, 75120 Uppsala, Sweden}%
\author{J\'an Rusz}%
    \affiliation{Department of Physics and Astronomy, Uppsala University, P.O. Box 516, 75120 Uppsala, Sweden}%

\date{\today}%

\begin{abstract}
We present a \highlight{computational method and apply it to study} phonon relaxation in face-centered cubic (fcc) nickel (Ni). 
The phonons are excited beyond their thermal equilibrium population, and the relaxation behavior is analyzed as a function of both the wave vector $\vec{q}$\, and the phonon frequency $\omega$. 
To efficiently investigate these excitations, we introduce a trajectory post-processing technique, the \textit{frozen-trajectory excitation}, which facilitates the $(\vec{q},\omega)$-resolved analysis.
Molecular dynamics simulations combined with frozen-phonon multislice calculations predict relaxation signatures observable with time-resolved transmission electron microscopy (TEM) at 10\textendash{}20\fs{} resolution. %
Our findings indicate mode dependence in the relaxation processes, highlighting the importance of considering phonon-specific behavior in ultrafast dynamics.
\end{abstract}

\maketitle

\section{Introduction}

\highlight{In 1996, Beaurepaire~\etal{} reported the discovery of ultrafast laser-induced demagnetization in face-centered cubic (fcc) nickel~\cite{Beaurepaire_ultrafast_1996}. They observed that a femtosecond laser pulse caused a rapid reduction in magnetization, within a few hundred femtoseconds, followed by a slower relaxation back to equilibrium. This discovery sparked significant interest because of its implications for subpicosecond magnetization switching, which holds great potential for future ultrafast magnetic storage technologies.}

\highlight{The excitation and subsequent de-excitation processes involve the intricate interplay between three coupled reservoirs\,\textemdash{}\,spins, the crystal lattice, and electrons\,\textemdash{}\,each characterized by distinct effective temperatures. To explain these dynamics, Beaurepaire~\etal{} introduced the so-called \textit{three-temperature model}~\cite{Beaurepaire_ultrafast_1996}, which has since become a leading tool for understanding this phenomenon. Extensions and refinements of this model have been proposed ~\cite{Koopmans_explaining_2010,Zahn_lattice_2021,Pankratova_heat-conserving_2022}, aiming to capture the complex interactions involved with greater accuracy.}

\highlight{Despite nearly three decades of research, the role of lattice dynamics in the ultrafast demagnetization process remains relatively underexplored, with most studies focusing on the spin and electron systems~\cite{Battiato_superdiffusive_2010,Roth_temperature_2012,cheng_ultrafast_2024, hui_attosecond_2024}. However, recent experimental and theoretical advancements have increasingly highlighted the critical contribution of lattice dynamics to this process~\cite{dettori_simulating_2017,Dornes_einstein-de-haas_2019,Maldonado_tracking_2020,Ritzmann_theory_2020,Tauchert_polarized_2022,barantani_ultrafast_2024}.}

\highlight{To contribute to bridging this gap, here we develop a computational scheme capable of isolating and capturing the role of inter-phonon interactions in this phenomenon. This is achieved by introducing a frequency- and phonon-wave-vector-resolved [($\vec{q},\omega$)-resolved] physical change in the simulated phonon mode population and capturing its evolution in time.}

\highlight{Complementing our computational efforts, we identify transmission electron microscopy (TEM) as a promising experimental tool for probing the ultrafast lattice dynamics. Advances in TEM technology, including aberration correction~\cite{Haider_electron_1998,Batson_sub-Angstrom_2002}, electron beam monochromation for resolving vibrational modes~\cite{Krivanek_vibrational_2014}, direct electron detectors~\cite{Li_electron-counting_2013,Hart_direct_2017}, and the incorporation of ultrafast laser pulses with fast cameras~\cite{Zewail_four-dimensional_2010,Feist_ultrafast_2017,Tauchert_polarized_2022,Gaida_attosecond_2024,Kim_high-resolution_2024,barantani_ultrafast_2024,hui_attosecond_2024}, enable atomic-scale spatial and temporal resolution. These developments suggest that the studies of ultrafast phonon dynamics using electron energy loss spectroscopy (EELS) are now within reach.}

\highlight{In this work, we focus on the computational study of ultrafast dynamics of phonons in fcc Ni, specifically their relaxation after excitation beyond thermal equilibrium. Using the \LAMMPS{} package~\cite{thompson_lammps_2022}, we introduce a method, referred to as \textit{frozen-trajectory excitation}, which enables the resolution of phonon excitations as a function of $\vec{q}$ and $\omega$. This method, designed for atomistic-scale molecular dynamics (MD) simulations, is flexible and efficient, with potential extensions to spin dynamics simulations~\cite{eriksson_atomistic_2017}.}
The implemented excitation introduction and the subsequent free relaxation over several picoseconds, along with the frozen-phonon multislice method~\cite{loane_thermal_1991}, employing \drprobe{}~\cite{barthel_dr_2018} for electron wave-function calculations, allow to analyze the evolution of resulting electron diffraction patterns in time.

\highlight{The theoretical framework presented here builds on the reciprocal space decomposition scheme employed in the frequency-resolved frozen-phonon multislice (FRFPMS) method~\cite{zeiger_efficient_2020,zeiger_frequencyresolved_2021}. Our approach extends this scheme to allow for arbitrary excitations and the reconstruction of complete molecular dynamics trajectories for subsequent free relaxation simulations. In an earlier study,  Dettori~\etal{} utilized a colored hotspot thermostat to selectively heat phonons within a specific energy range, achieving temperatures well above equilibrium~\cite{dettori_simulating_2017}. However, their technique does not allow for wave-vector-dependent excitations.}

\section{Method}\label{sec:method}
\subsection{Overwiev}



\begin{figure}[ht!]
\center
    \fhighlight{\begin{tikzpicture}[rounded/.style={rectangle, draw=black, rounded corners, align=center, minimum width=0.9\columnwidth},empty/.style={align=center}]
    \node[empty] at (0,0) (label1) {(I) Equilibrium trajectory generation};
    
    \begin{scope}[on background layer]
        \node[rounded, fit=(label1), fill=red!10] (step1) {};
    \end{scope}
    
    \node[empty] at (0,-2) (label3) {(II) Frozen trajectory excitation:};

    \node[rounded, fill=white] at (0,-2.75) (node5) {FFT'd displacements in ($\vec{q},\omega$)-space};
    \node[rounded, fill=white, fill=white] at (0,-4) (node6) {Displacements of the excited system};
    \node[rounded, fill=white] at (0,-5.25) (node7) {Full trajectory with velocities};

    \draw[->] (node5) -- (node6) node[empty,midway,right] {Excitation};
    \draw[->] (node6) -- (node7) node[empty,midway,right] {Trajectory reconstruction};

    \begin{scope}[on background layer]
        \node[rounded,fit=(label3)(node5)(node6)(node7), fill=green!10] (step2) {};
    \end{scope}
    
    \node[empty] at (0,-7.5) (label7) {(III) Free relaxation.\\ Multiple relaxation runs are spawned\\ from a single excited trajectory};
    
    \begin{scope}[on background layer]
        \node[rounded,fit=(label7), fill=blue!10] (step3) {};
    \end{scope}

    \draw[->] (step1) -- (step2) node[empty,midway,left] {Save atomic positions\\ from LAMMPS run} node[empty,midway,right] {Remap atom indices\\ to primitive basis};
    \draw[->] (step2) -- (step3) node[empty,midway,right] {load selected snapshot\\ to \LAMMPS{}} node[empty,midway,left] {Remap atom indices\\ back to \LAMMPS{}};
    \end{tikzpicture}}
\caption{\label{fig:schematic} 
\highlight{Overview of the \textit{frozen trajectory excitation} protocol, presenting steps performed to obtain atomic configurations at different time delays during the relaxation process.}
}
\end{figure}

\highlight{Our calculations consist of three main phases, (I)\textendash{}(III), summarized in Fig.~\ref{fig:schematic}, that were performed using an in-house software designed to integrate \LAMMPS{} via its ``as a library'' feature~\cite{thompson_lammps_2022}.
In the first phase, (I), the system is created and equilibrated using  \LAMMPS{} subroutines. 
Subsequently, we obtain the time-dependent position of all atoms\,\textemdash{}\,the MD \textit{trajectory}\,\textemdash{}\,during the equilibrium finite-temperature simulations of the system (from \LAMMPS{}).
Next, in the second phase, (II), this trajectory is manipulated using the \fftw{} fast Fourier transform (FFT) library~\cite{frigo_design_2005}, following the procedure detailed below, to introduce the desired excitation. 
We call the procedure a frozen trajectory excitation (FTE).
In the final phase, (III), we allow for a free MD relaxation of the excited system and calculate time-resolved electron diffraction patterns.}

\highlight{In the following subsections~\ref{sec:trajectory-generation}\textendash{}\ref{sec:relaxation}, we provide an underlying physics background for the method.
Sec.~\ref{sec:calculations-details} provides details about simulation parameters, whereas App.~\ref{app:details} contains a comprehensive description of implementation details.}


\subsection{\label{sec:trajectory-generation}Equilibrium trajectory generation}

\highlight{In the first stage\,\textemdash{}\,(I) Equilibrium trajectory generation\,\textemdash{}\,we save complete sets of atomic positions in the system (referred to later as snapshots) at regular time intervals, during an ordinary MD run.
It is preceded by proper equilibration and thermalization, and the length of this MD run is matched to allow the desired frequency resolution to be obtained in subsequent stages, incorporating FFT.}
This is then repeated several times with new random seeds, as described further.
%
%
Each atom's trajectory \highlight{is} decomposed into three components: the origin of the unit cell $\vec{r}^{\,0}$, the basis site position $\vec{b}$, and the atomic displacement $\vec{u}$. Explicitly,

\begin{equation}
\label{eqn:trajectory_components}
    \vec{r}_{h, k, l, i}(t) = \vec{r\,}^0_{h,k,l} + \vec{b}_i + \vec{u}_{h, k, l, i}(t),
\end{equation}
where $i$ is the unit cell basis \highlight{site} index, and $h$, $k$, and $l$ are \highlight{basis translation indices} within the supercell, corresponding to integer multiples of the lattice vectors $\vec{a}_1$, $\vec{a}_2$, and $\vec{a}_3$, i.e. $\vec{r\,}^0_{h,k,l} = h\vec{a}_1 + k\vec{a}_2 + l\vec{a}_3$.
The displacements $\vec{u}$ can thus be represented in a \highlight{6}-dimensional hypercube. \highlight{In our approach, we consider each basis site separately, and \textit{x}, \textit{y}, and \textit{z} displacement components are treated equivalently. Hence, we reduce the basis site and cartesian coordinate dimensions, obtaining a 4-dimensional dataset for further treatment, where the FFT indexed by \textit{h}, \textit{k}, and \textit{l} yields phonon wave vector ($\vec{q}$) spectrum, and the FFT indexed by timeframe corresponds to the frequency spectrum.}

\subsection{\label{sec:excitation}Frozen-trajectory excitation}

\highlight{Via the aforementioned FFT, we calculate $\vec{u}_i(\vec{q},\omega)$\,\textemdash{}\,in principle for each primitive basis atom, which in the case of fcc Ni reduces to one atom only ($i = 1$).
Subsequently, we introduce an excitation into the system by applying a band-enhance filter $F(\vec{q},\omega)$, designed to amplify an arbitarily chosen region of the $(\vec{q},\omega)$-space (see Appendix for details).
Creating such a modified displacement field $\vec{u}'_i(\vec{q},\omega)$, corresponding to an excited state, allows for treating the resulting trajectory as if it has constantly been in an out-of-equilibrium state in which a specific range of phonon modes is thermally overpopulated.}



\highlight{In order for the obtained trajectory to be useful as a starting point for subsequent simulations, we require a full set of coordinates $\vec{R} = \{\{\vec{r}_i', \vec{v}_i'\}\}_{i=1}^{N_{\rm atom}}$, where both atomic positions and velocities are included. 
The velocities can be obtained from $\vec{u}'(\vec{q}, \omega)$ by calculating the time derivative of displacements that, in energy space, gives:}
\begin{equation}
\label{eqn:x_to_v_transformation}
        \vec{v}'(\vec{q}, \omega) = -i \omega \vec{u}'(\vec{q}, \omega).
\end{equation}
\highlight{We obtain $\vec{v}'(h, k, l, t)$ and $\vec{u}'(h, k, l, t)$ through inverse Fourier transforms, and from there finally obtain atomic positions in the excited trajectory.}


\subsection{\label{sec:relaxation}Free relaxation and diffraction patterns calculation}


\highlight{From the excited trajectory, we select a subset of $N_{\rm S}$ snapshots spaced sufficiently apart to minimize the correlation between them.
%
%
Using each of those snapshots, we perform a simulation of the relaxation process, obtaining atomic configurations $\vec{R}(\Delta t) = \{\vec{r}_i(\Delta t)\}_{i=1}^{N_{\rm atom}}$ at specific time delays $\Delta t$.
At each time delay, we compute the corresponding TEM diffraction intensities\,\textemdash{}\,that, in general, can depend on the in-plane momentum transfer $\vec{q}_{\perp}$, the electron beam position $\vec{r}_b$, and the atomic configuration $\vec{R}$.
This is done employing the frozen-phonon multislice (FPMS) approach~\cite{loane_thermal_1991}, which is equivalent to the quantum excitation of phonons (QEP) framework~\cite{forbes_quantum_2010,lugg_atomic_2015}.
The so-called multislice auxiliary wavefunction $\psi(\vec{q}_{\perp}, \vec{r}_b, \vec{R}(\Delta t))$ is obtained using \drprobe{}~\cite{barthel_dr_2018}.
Explicitly, the intensities are calculated as:
\begin{align}
    \label{eqn:qep-1} I_{\rm incoh}(\vec{q}_{\perp}, \vec{r}_b, \Delta t) &= \frac{1}{N_S}\sum\limits_{j=1}^{N_S} \lvert\psi(\vec{q}_{\perp}, \vec{r}_b, \vec{R}_j(\Delta t))\rvert^2 \\
    \label{eqn:qep-2} I_{\rm coh}(\vec{q}_{\perp}, \vec{r}_b, \Delta t) &= \left\lvert\frac{1}{N_S}\sum\limits_{j=1}^{N_S} \psi(\vec{q}_{\perp}, \vec{r}_b, \vec{R}_j(\Delta t))\right\rvert^2 \\
    \label{eqn:qep-3} I_{\rm vib}(\vec{q}_{\perp}, \vec{r}_b, \Delta t) &= I_{\rm incoh}(\vec{q}_{\perp}, \vec{r}_b, \Delta t) - I_{\rm coh}(\vec{q}_{\perp}, \vec{r}_b, \Delta t),
\end{align}
where $I_{\rm incoh}$, $I_{\rm coh}$, and $I_{\rm vib}$ represent respectively the total scattering intensity (corresponding to the incoherent average), the elastic intensity (coherent average), and the vibrational intensity (thermal diffuse scattering\,\textemdash{}\,TDS).}


\highlight{Note that the averaging is performed over snapshots $\vec{R}_j(\Delta t)$ from the $N_{\rm S}$ different realizations of the relaxation process. This differs from usual applications of FPMS utilizing MD simulations, where snapshots originate from the same trajectory~\cite{lofgren_influence_2016,aveyard_modeling_2014,muller_simulation_2001,chen_comparison_2023}.
It is worth emphasizing that such extension of the standard QEP (and FPMS) method provides a time-dependent description of the thermal diffuse scattering, i.e., the inelastic phonon scattering in TEM.
}

\section{\label{sec:calculations-details}Calculation details}



\moved{Most molecular dynamics (MD) simulations were conducted with integration schemes based on the time-reversible measure-preserving Verlet and rRESPA integrators by Tuckerman~\etal{}~\cite{tuckerman_liouvilleoperator_2006}. 
In the free MD simulations, we employed the pure velocity-Verlet integrator. 
A consistent time step of 1\fs{} was used throughout all simulations.}


\moved{We considered a 28~$\times$~28~$\times$~28 supercell of the conventional 4-atom \highlight{face-centered cubic} (fcc) Ni cubic unit cells oriented along the (001) crystal direction. 
%
%
The lattice parameter was optimized using the isothermal-isobaric (NpT) ensemble, resulting in lattice parameter $a = 3.5329$~\AA{} at 300 K.
\highlight{It yielded an approximately 10~nm~$\cross$~10~nm~$\cross$~10~nm system, repeated with periodic boundary conditions in all spatial dimensions.}
%
%
With the optimized $a$, the calculations were restarted in the canonical (NVT) ensemble, and the initial atomic velocities were randomized to match the target temperature. 
The system was then equilibrated for 2\ps{} at the target temperature of 300 K. 
Both the NpT and NVT stages employed the Nos\'{e}-Hoover thermostat and barostat~\cite{parrinello_polymorphic_1981}, with the equations of motion based on the implementation by Shinoda~\etal{}~\cite{shinoda_rapid_2004}.
%
%
After the equilibration, we performed simulations spanning 1 ps within the canonical (NVT) ensemble, saving snapshots of the system at regular time intervals of 20~\fs{}.
%
%
To ensure computational efficiency and cost-effectiveness, we described the interatomic interactions with the spectral neighbor analysis pattern machine-learning interatomic potential (SNAP ML-IAP~\cite{thompson_spectral_2015}, referred to further as SNAP) in the parametrization of Zuo~\etal{}~\cite{zuo_performance_2020}.}

\highlight{The anharmonicity of the potential allows for phonon-phonon scattering and results in finite lifetimes of phonon excitations. 
The employed potential is trained on parameters derived from a fully quantum treatment of 800 distorted crystal structures. 
This training includes the redistribution of electron charge density. 
As a result, the approach ensures a near-DFT level of accuracy while also implicitly accounting for the partial influence of the electronic cloud on atomic vibrations. 
However, it is important to note that fully incorporating electron-phonon interaction would necessitate either quantum mechanical molecular dynamics simulations or the pre-training of a custom interatomic potential. 
This pre-training would require a dataset that explores a vast portion of the fcc Ni configurational space, particularly focusing on heavily distorted structures.}

%
\highlight{Next, we introduced excitations to the trajectory, using the FTE approach, described in the previous section.}
\moved{\highlight{We increased 20-fold the FFT amplitudes of a ($\vec{q},\omega$) region corresponding to} transverse acoustic (TA) phonons separately at four distinct points in the $q_z=0$ plane of the reciprocal space.
These points include the high-symmetry X point in both the $\hat{q}_y$ and $\hat{q}_z$ directions, as well as two intermediate points along the $\Delta$ high-symmetry line. 
The selected excitation leads to a temperature increase in the system by approximately 25\,K due to the finite size of the system.}


\highlight{We chose seven snapshots from a middle section of one modified trajectory as initial frames for further separate, free relaxation in the microcanonical (NVE) ensemble. 
We repeated the procedure 7 or 14 times depending on the specific simulation scheme, resulting in $N_{\rm S} = 49$ or $98$ uncorrelated relaxation realizations, respectively. 
In the initial stage of the study, 100 snapshots were saved during the system's relaxation at 10~\fs{} intervals. 
In the later stages, the sampling count was increased to 200 frames of the system's relaxation, with the time interval extended to 20~\fs{}.
For each of the relaxation runs and for each time frame in those runs, we calculate TEM diffractograms with a parallel incident electron beam\,\textemdash{}\,to avoid complications with interpretation due to the convergent electron beam\,\textemdash{}\,at an acceleration voltage of 100~kV.}

\section{Results and discussion}

\highlight{\subsection{Initial results}}


\begin{figure}[ht!]
\center
\fmoved{\includegraphics[width=0.95\columnwidth]{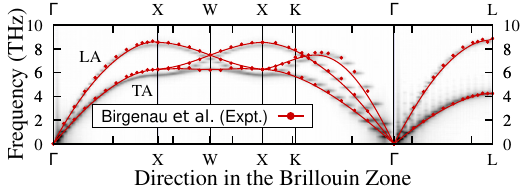}}
\caption{\label{fig:phonon_dispersion_equilibrium}%
\moved{Comparison of the \highlight{calculated phonon FFT intensity} in fcc Ni \highlight{(grayscale)} with experimental phonon dispersion from Birgenau~\etal{}~\cite{birgeneau_normal_1964}. The intensity has been calculated from atomic displacements according to Eq.~\ref{eqn:phonon_dispersion_equation} \highlight{and subsequently normalized. Longitudinal acoustic (LA) and transverse acoustic (TA) phonon branches are labelled for further reference.}}
}
\end{figure}

\moved{To evaluate the accuracy of the simulations, we compared the phonon dispersion stemming from the chosen approach with experimental results. 
Typically, phonon dispersions are obtained from MD trajectories via a Fourier transform $\mathcal{F}$ of the velocity autocorrelation function~\cite{lee_initio_1993}. 
However, we approximate phonon dispersions~$\omega_{\rm ph}(\vec{q})$ \highlight{as pronounced local maxima of} the Fourier-transformed atomic displacements $I(\vec{q}, \omega)$: 
\begin{equation}
\label{eqn:phonon_dispersion_equation}
    \omega_{\rm ph}(\vec{q}) = {\rm max}(I(\vec{q}, \omega)) = {\rm max}(\lvert \mathcal{F}(\vec{u}(\vec{r}^{\,0}, t)) \rvert).
\end{equation}
This approach is computationally equivalent to using velocity autocorrelation, up to a constant~\cite{carreras_dynaphopy_2017}.
Note that the authors in Ref.~\cite{carreras_dynaphopy_2017} worked with velocities, which are phase-shifted in relation to the displacements. 
We show this quantity along with the experimental results by Birgenau~\etal{}~\cite{birgeneau_normal_1964} in Fig.~\ref{fig:phonon_dispersion_equilibrium}. 
%
%
The phonon dispersion calculated from our simulations shows good agreement with the experimental results, showing that we achieve accuracy comparable to density functional theory (DFT) calculations at much reduced computational costs.}


\begin{figure*}[ht!]
\center
\fmoved{\includegraphics[width=0.9\textwidth]{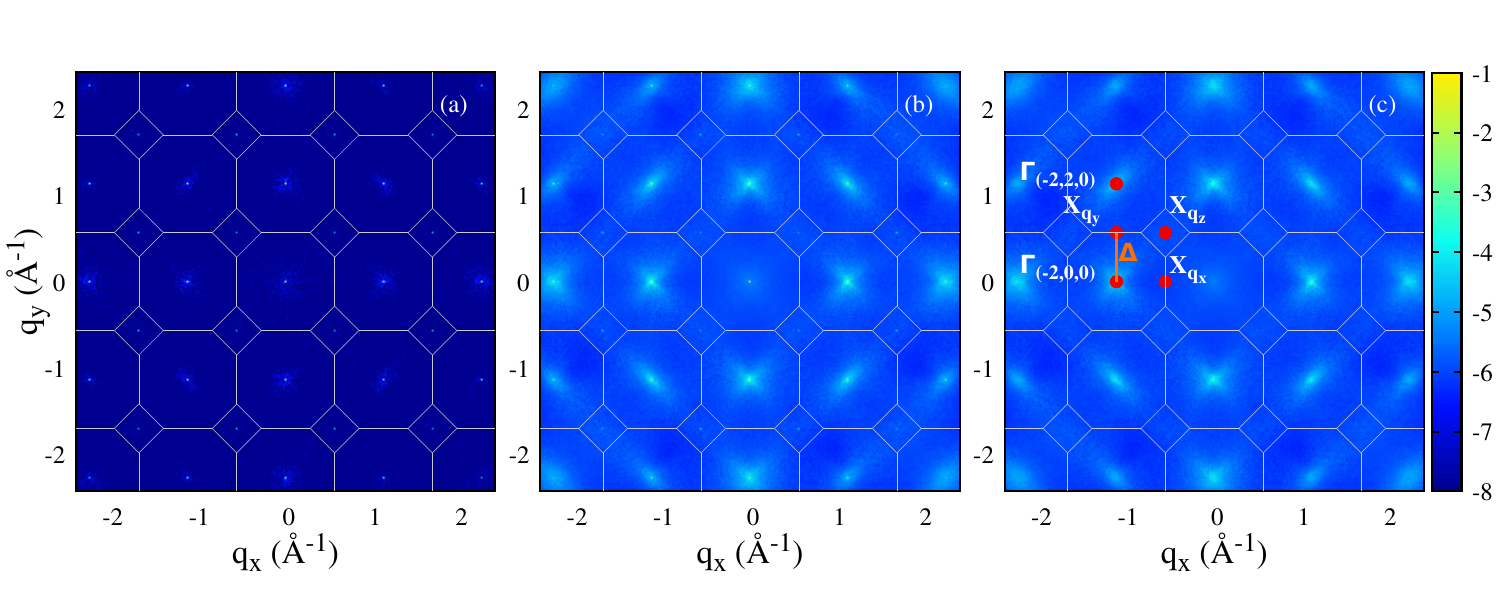}}
\caption{\label{fig:diffractograms_equilibrium}%
\moved{Coherent (a) and incoherent (b) averages of exit wavefunctions over 350 atomic configurations in simulated electron diffraction, using a parallel beam incident on an fcc Ni supercell (28 $\times$ 28 $\times$ 28 conventional 4-atom unit cells) in thermodynamic equilibrium at 325~K. 
Panel (c) shows the thermal diffuse (inelastic) scattering diffraction pattern.
The color scale indicating fractional intensities is logarithmic.
The thin lines show the first Brillouin Zone boundaries in the diffraction plane.}
}
\end{figure*}

\moved{Figure~\ref{fig:diffractograms_equilibrium} shows the calculated electron diffraction pattern contributions in the system without the induced excitation, including the coherent (a), incoherent (b), and vibrational (c) components, based on $N=350$ snapshots sampled from an equilibrium NVT trajectory before excitation and NVE relaxation (see Eqs.~(\ref{eqn:qep-1})\textendash{}(\ref{eqn:qep-3})). 
In Fig.~\ref{fig:diffractograms_equilibrium}(a), minor noise is visible around the Bragg spots. 
This noise results from incomplete sampling of the system's configurational space and should decrease with increasing statistics. 
Importantly, the noise is approximately two orders of magnitude weaker than the vibrational intensity in these regions, as shown in Fig.~\ref{fig:diffractograms_equilibrium}(c). 
Additionally, kinematically forbidden reflections can be observed. 
These occur because the distribution of atoms in individual atomic planes differs from that of atoms in a $z$-projected unit cell. 
Beam propagation effects cause imperfect destructive interference, leading to the appearance of these kinematically forbidden Bragg spots, albeit at low intensity.}

\highlight{\subsection{Excited system}}


\begin{figure}[ht!]
\center
\includegraphics[width=0.95\columnwidth]{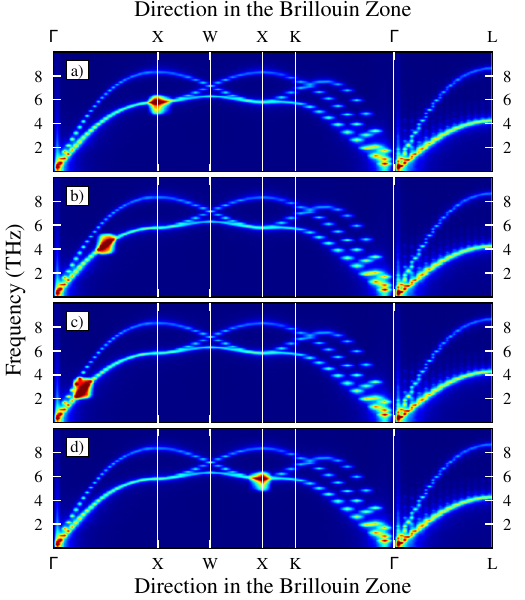}
\caption{\label{fig:phonon_dispersion_excited}%
Phonon \highlight{intensity} in fcc Ni after applying the simulated excitation filter. 
Panels (a) and (d) show excitations \highlight{A and D}, at X$_{\hat{q}_y}$ and X$_{\hat{q}_z}$ points. 
Panels (b) and (c) depict excitations \highlight{B and C}, in selected points along the $\Delta$ high-symmetry line in the $\hat{q}_y$ direction. Normalized intensities are reflected in the heatmap color scheme.
}
\end{figure}

\begin{figure}[ht!]
\center
\includegraphics[width=0.95\columnwidth]{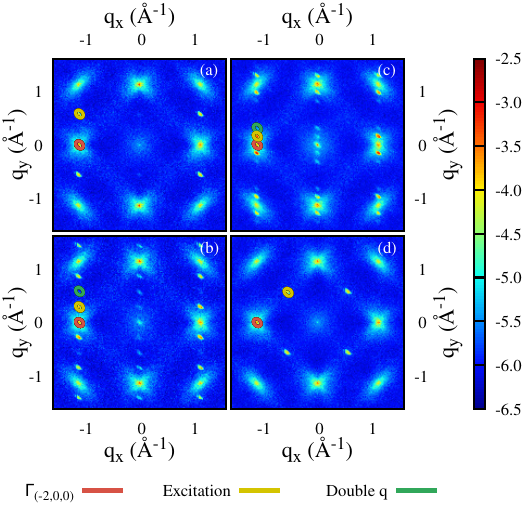}
\caption{\label{fig:diffractograms_excited}%
TDS diffraction patterns of the excited system before relaxation: 
(a) Excitation A, at the X$_{\hat{q}_y}$ point, 
(b) Excitation B, along the $\Delta$ high-symmetry line in the $\hat{q}_y$ direction, 
(c) Excitation C, along the $\Delta$ high-symmetry line in the $\hat{q}_y$ direction, and 
(d) Excitation D, at the X$_{\hat{q}_z}$ point. 
The colored ellipses indicate the integration areas for intensity plots, which are presented in subsequent figures with corresponding colored lines.
Intensities in all panels are presented on a logarithmic scale.
}
\end{figure}

Figure~\ref{fig:phonon_dispersion_excited} presents the ($\vec{q}, \omega$)-resolved oscillation amplitudes after excitations have been introduced.
So far, we have not explicitly discussed the rationale behind the selection of the excitation points. 
 X was chosen because it is the only high-symmetry point for which the reflections of all its symmetricaly equivalent copies\highlight{\,\textemdash{}\,X$_{\hat{q}_x}$ (X high symmetry point in the $\hat{q}_x$ direction), X$_{\hat{q}_y}$ (X high symmetry point in the $\hat{q}_y$ direction), and X$_{\hat{q}_z}$) (X high symmetry point in the $\hat{q}_z$ direction\,\textemdash{}\,}can be observed in the diffraction pattern in the selected simulation conditions, see Fig.~\ref{fig:diffractograms_equilibrium}(c). 
Additionally, excitation points along the $\Delta$ high-symmetry line in the $\hat{q}_y$ direction\,\textemdash{}\,specifically, halfway between the $\Gamma$ and X high-symmetry points, and at half the frequency of the X point\,\textemdash{}\,enable us to observe multi-phonon scattering events during the relaxation process, as we will show further.
\highlight{To simplify following descriptions and provide a more intuitive link between text and figures, we henceforth label the excitations mentioned earlier as:
\begin{itemize}
    \item A\,\textemdash{}\,excitation in X$_{\hat{q}_y}$ (Fig.~\ref{fig:phonon_dispersion_excited}(a)),
    \item B\,\textemdash{}\,halfway between the $\Gamma$ and X$_{\hat{q}_y}$ high-symmetry points (Fig.~\ref{fig:phonon_dispersion_excited}(b)), 
    \item C\,\textemdash{}\,along the $\Delta$ high symmetry line towards X$_{\hat{q}_y}$ and at half the TA mode frequency of the X$_{\hat{q}_y}$ point (Fig.~\ref{fig:phonon_dispersion_excited}(c)), 
    \item D\,\textemdash{}\,excitation in X$_{\hat{q}_z}$ (Fig.~\ref{fig:phonon_dispersion_excited}(d)).\end{itemize}
For convenience, these A\textendash{}D labels also correspond to a\textendash{}d subfigures in Figs.~\ref{fig:diffractograms_excited}\textendash{}\ref{fig:relaxation_intensities_scans_3d}.
The introduced excitation in all cases resulted in raising the temperature of the system to approximately 325 K.}


Figure~\ref{fig:diffractograms_excited} presents the diffraction patterns of the TDS \highlight{right after the excitation}.
The \highlight{changes} appear as intensity peaks at their corresponding $\vec{q}_\perp = (q_x,q_y)$ values. 
In panels \ref{fig:diffractograms_excited}(b) and \ref{fig:diffractograms_excited}(c), additional reflections are observed at double the $\vec{q}$ of the induced excitation (highlighted by red ellipses). 
In the initial frame (0$^{th}$ frame), we can confidently conclude that this effect is not caused by a higher-than-thermal excitation of phonon modes with double momentum transfer, as the phonon dispersion and population statistics are well-resolved. 
Therefore, these additional local intensity maxima can only be attributed to multi-phonon scattering processes.


The excitation peaks in the diffractograms are all elongated in the $(-1,1)$ direction (i.e. at $135\degree$ from the $q_x$ axis). 
This results from the cubic filter ($\Delta q_{a_{1}} = \Delta q_{a_{2}} = \Delta q_{a_{3}}$) applied in the primitive coordinate system:
\begin{align*}
\label{eqn:primitive_vectors}
    \vec{q}_{a_{1}} &= (\hspace{\minuslength}a^{-1}, -a^{-1}, \hspace{\minuslength}a^{-1}), \\
    \vec{q}_{a_{2}} &= (\hspace{\minuslength}a^{-1}, \hspace{\minuslength}a^{-1}, -a^{-1}), \\
    \vec{q}_{a_{3}} &= (-a^{-1}, \hspace{\minuslength}a^{-1}, \hspace{\minuslength}a^{-1}).
\end{align*}
When described in Cartesian coordinates, this cubic filter transforms into a rhombohedron with its short diagonal aligned along the (111) direction in reciprocal space. 
The projection of this rhombohedron onto the (001) plane in the $\vec{q}$-space specified in the cartesian basis produces the skew observed in all panels of Fig.~\ref{fig:diffractograms_excited}.

\begin{figure}[ht!]
\center
\fhighlight{\includegraphics[width=0.95\columnwidth]{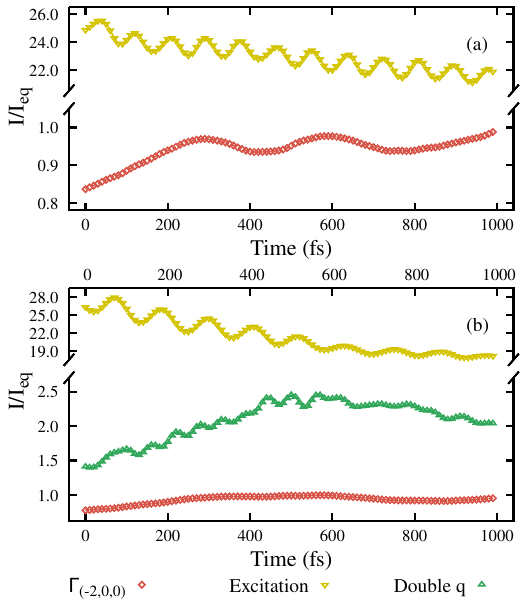}}
\caption{\label{fig:relaxation_intensities_1ps_dense}%
Integrated TDS intensities over a 1~\ps{} trajectory with a sampling period of 10~\fs{}, relative to the intensities in thermal equilibrium at 325 K. 
Panel (a) shows the relaxation of the excitation A, at the X$_{\hat{q}_y}$ point. 
Panel (b) depicts the relaxation of the excitation B, along the $\Delta$ high-symmetry line in the $\hat{q}_y$ direction. 
Panel (b) reveals a possible energy transfer to the X-point phonons.
The plots represent the relaxation processes of the excitations shown in Fig.~\ref{fig:diffractograms_excited}, with colors and panel order corresponding to that figure.
}
\end{figure}
\begin{figure*}[ht!]
\center
\fhighlight{\includegraphics[width=0.95\textwidth]{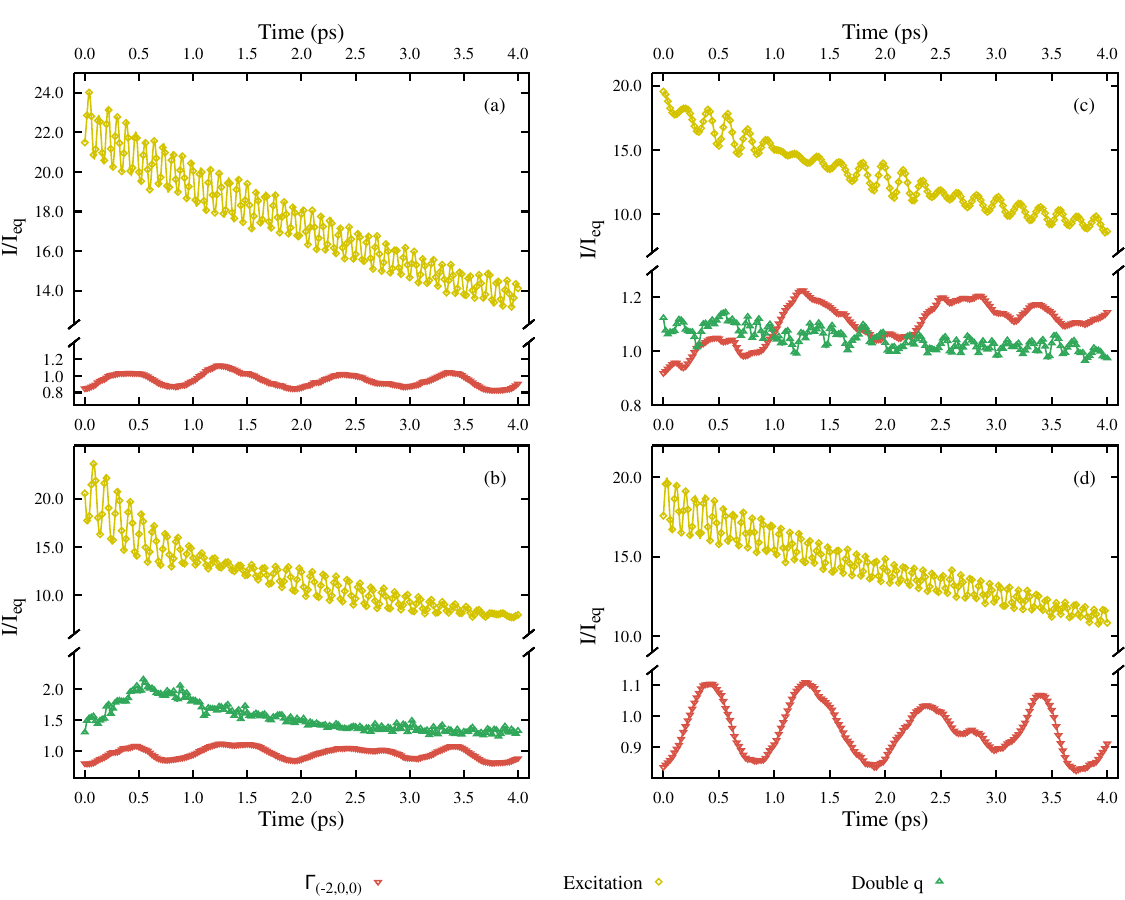}}
\caption{\label{fig:relaxation_intensities_4ps_sparse}%
Integrated TDS diffraction intensities, relative to the intensities in thermal equilibrium at 325 K, over a longer, 4~\ps{}, trajectory with a sampling period of 20~\fs{}. 
Panel (a) shows the relaxation of the excitation A, panel (b) depicts the relaxation of the excitation B, panel (c) presents the relaxation of the excitation C, and panel (d) shows the relaxation of the excitation D. 
The plots correspond to the relaxation processes of the excitations shown in Fig.~\ref{fig:diffractograms_excited}, with matching colors and panel order.
}
\end{figure*}
\begin{figure*}[ht!]
\center
\includegraphics[width=0.95\textwidth]{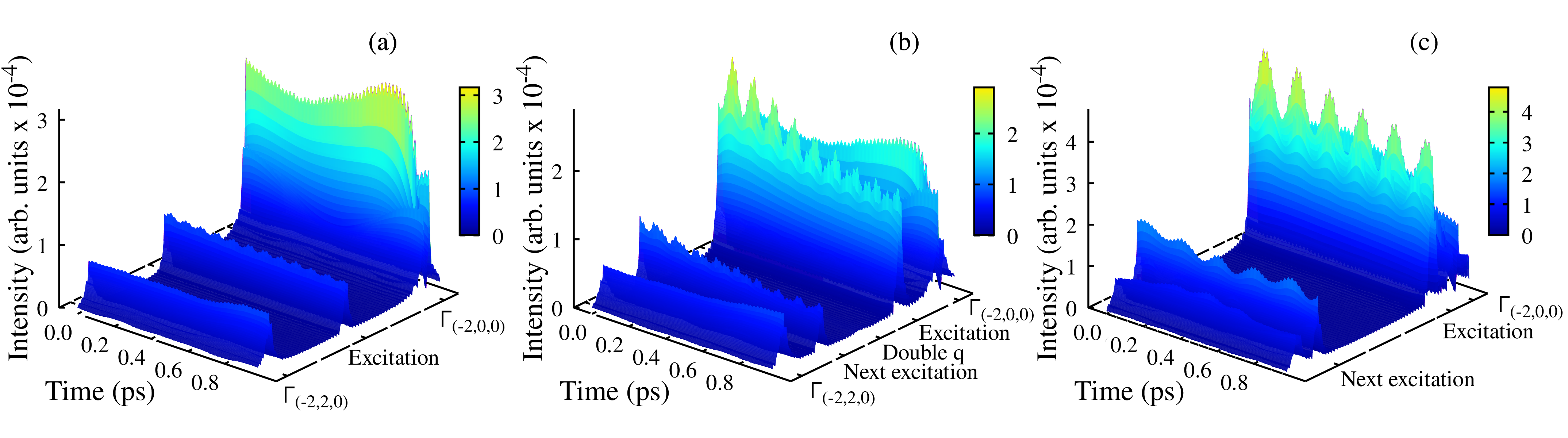}
\caption{\label{fig:relaxation_intensities_scans_3d}%
TDS diffraction intensities along the $\Delta$ ($\Gamma$\textendash{}X) high symmetry line, presenting relaxation of: (a) excitation A (in X$_{\hat{q}_y}$ point, (b) excitation B, along the $\Delta$ high symmetry line in $\hat{q}_y$ direction, and (c) excitation C, also along the $\Delta$ high symmetry line in $\hat{q}_y$ direction. 
Plots present the relaxation process of excitations presented in Fig.~\ref{fig:diffractograms_excited}, considering the full unit-cell-resolved $\Delta$-line scan.
Same data\,\textemdash{}\,integrated over a finite aperture surrounding selected points\,\textemdash{}\,is presented in more detail in Figs.~\ref{fig:relaxation_intensities_1ps_dense} and~\ref{fig:relaxation_intensities_4ps_sparse}.
All plots use the respective order of panels.
}
\end{figure*}

Therefore, for further analysis, we integrate the vibrational intensities over an elliptic aperture $\Omega$: 
\begin{equation}
\label{eqn:integration}
    I_{\rm vib}[\Omega(\vec{q}_\perp), \vec{r}_b] = \iint\limits_{\Omega(\vec{q}_\perp)} I_{\rm vib}(\vec{q}_{\perp}, \vec{r}_b) \diff \vec{q}_{\perp},
\end{equation}
The ellipses are shown in Fig.~\ref{fig:diffractograms_excited}, with colors corresponding to the following intensity-versus-time plots in Figs.~\ref{fig:relaxation_intensities_1ps_dense} and~\ref{fig:relaxation_intensities_4ps_sparse}.

\highlight{\subsection{Relaxation\texorpdfstring{\,\textemdash{}\,}{}shorter timescale, higher temporal resolution}}


Fig.~\ref{fig:relaxation_intensities_1ps_dense} presents the first 1\ps{} of the relaxation process of two of the selected excitations \highlight{(A and B)} with a temporal resolution of 10\fs{}.
The diffractograms are averaged over 98 uncorrelated relaxation trajectories realized according to the procedure described in Sec.~\ref{sec:relaxation}.
\highlight{All panels show TDS intensities related to equilibrium TDS at 325 K.}
There are two key observations. 

Firstly, the excitation \highlight{B (along the $\Delta$ line)} (Fig.~\ref{fig:relaxation_intensities_1ps_dense}(b)) relaxes significantly faster than the excitation \highlight{A (in X$_{\hat{q}_y}$ point)} (Fig.~\ref{fig:relaxation_intensities_1ps_dense}(a)). 
\highlight{In Fig.~\ref{fig:relaxation_intensities_1ps_dense}(b), the yellow curve drops below 70\% of its initial value within the first picosecond, while the corresponding excitation in Fig.~\ref{fig:relaxation_intensities_1ps_dense}(a) is still over 85\% of the initial intensity after the same period.}
Interestingly, this faster relaxation is accompanied by an intensity enhancement at double the excitation point's $\vec{q}$ (see the green curve in Fig.~\ref{fig:relaxation_intensities_1ps_dense}(b)). 
This may indicate ongoing energy transfer to phonon modes associated with the X-point. 
This behavior differs from the multi-phonon excitations seen in Fig.~\ref{fig:diffractograms_excited}, as scattering from the excited modes decreases over time, and the two-phonon scattering should also diminish. 
However, a mild but distinct intensity increase is observed up to around 600\ps{} \highlight{after the excitation}.
%
%
This \highlight{double-$\vec{q}$} intensity enhancement \highlight{visible in Fig.~\ref{fig:relaxation_intensities_1ps_dense}(b) indicates} a more complex scenario, as it may reflect energy transfer, multi-phonon excitation at a specific frequency, or both.
%
%
\highlight{The fact that the multi-phonon scattering signal should decrease whenever the excitation itself decreases, leads to} a conclusion that the relaxation time constant $\tau$ is not only strongly dependent on $(\vec{q},\omega)$, but is also governed by intricate phonon-phonon interactions \highlight{including the energy transfer between modes}. 
We, \highlight{hence}, confirm the mode-dependence in the relaxation processes~\cite{Ritzmann_theory_2020} and the involvement of phonon-phonon interactions in the process by the real-time, explicit atomistic simulations.
This indicates that, in addition to electron-phonon and phonon-magnon interactions, phonon-phonon interactions must also be considered when analyzing the system’s magnetization dynamics. Regardless, the strong $(\vec{q},\omega)$-dependence of $\tau$ \highlight{emerging from explicit all-atom simulations} is clearly observed.


%
\moved{Secondly, we observe oscillations in the vibrational intensity in the excitation point, occurring at double the frequency of the excited phonon mode. 
Those are reminiscent of the real space atomic movements and stem from the incompleteness of configuration space sampling.
Since observed intensities come from squared electron-beam exit wave-functions, the \highlight{measured} frequency of oscillations is\highlight{\,\textemdash{}\,by trigonometric identities\,\textemdash{}\,}twice the frequency of phonon modes observable at given scattering angles.
It is noteworthy that similar intensity oscillations have been previously reported at the $\Gamma$-point in bismuth~\cite{qi_breaking_2020,sokolowski-tinten_femtosecond_2003,harmand_achieving_2013,fritz_ultrafast_2007,zhao_noninvasive_2021}.
Two factors currently limit the experimental observation of the behavior we describe in this work. 
First, to the best of our knowledge, there is no method for inducing an arbitrary $\vec{q}$-resolved phonon excitation in scanning transmission electron microscopy. 
Second, current hardware capabilities are only now reaching the femtosecond temporal resolution necessary to observe these effects~\cite{qi_breaking_2020,hui_attosecond_2024}.}

\highlight{\subsection{Relaxation\texorpdfstring{\,\textemdash{}\,}{}longer timescale, lower temporal resolution}}


Since the excitations presented in Fig.~\ref{fig:relaxation_intensities_1ps_dense} after 1\ps{} have not yet \highlight{fully} equilibrated, we repeated the simulations for a longer but less detailed trajectory spanning 4\ps{}, with a sampling period of 20\fs{}.
We also expanded the range of excited points in $(\vec{q}, \omega)$ space by introducing \highlight{one more excitation along the $\Delta$ line [C point, Fig.~\ref{fig:relaxation_intensities_4ps_sparse}(c)] and one more excitation in X point image [D point, Fig.~\ref{fig:relaxation_intensities_4ps_sparse}(d)]}. 
These calculations also served to cross-check the results shown in Fig.~\ref{fig:relaxation_intensities_1ps_dense}. The results, presented in Fig.~\ref{fig:relaxation_intensities_4ps_sparse}, are averaged over 49 new uncorrelated trajectories with different random seeds, with excitations introduced as described in Sec.~\ref{sec:excitation}.
Over this extended period, the difference in relaxation timescales between X-point excitations [Fig.~\ref{fig:relaxation_intensities_4ps_sparse}(a,d)] and selected excitations along the $\Delta$ high-symmetry line [Fig.~\ref{fig:relaxation_intensities_4ps_sparse}(b,c)] becomes even more pronounced.
Fig.~\ref{fig:relaxation_intensities_scans_3d} shows a 3D view of this evolution of the $\Delta$ scan intensity in time and depicts
the true magnitude of oscillations relative to the excitation amplitude.


Additionally, Figs.~\ref{fig:relaxation_intensities_4ps_sparse}(b) and~\ref{fig:relaxation_intensities_4ps_sparse}(c) reveal a beat in the vibrational intensity oscillations at the excitation point. 
This beat arises due to the finite size of both the excitation filter and the aperture used for integrating the intensity. 
The effect is more pronounced in Fig.~\ref{fig:relaxation_intensities_4ps_sparse}(c) because the excitation region clearly captures the longitudinal acoustic mode at the same $\vec{q}$, but in different $\omega$, resulting in the effective summation of vibrational intensities from two distinct phonon modes. 
Nevertheless, Fig.~\ref{fig:relaxation_intensities_4ps_sparse}(b) also shows that the enhancement affects a range of phonon frequencies, which are integrated over the aperture\highlight{, leading to similar, albeit less pronounced beat}. 
Consequently, this beat should disappear for a point excitation (or point aperture).
As discussed before, the oscillations stem from the atomic movement, so they should become negligible with sufficiently large sample averaging.


After 4\ps{}, excitations B and C along the $\Delta$ high-symmetry line [Fig.~\ref{fig:relaxation_intensities_scans_3d}(b, c)]
reach their half-lives due to interactions between multiple excited phonons, suggesting a relaxation time of the order of 4-5~ps, which is an order of magnitude well consistent with the literature \cite{Maldonado_tracking_2020,Pankratova_heat-conserving_2022}. 
In contrast, the intensity of excitations A and D (at X$_{\hat{q}_y}$ and at X$_{\hat{q}_z}$) decrease by only around 30\,\textemdash{}\,35\% over the same period. 


The $\tau$ derived from the three-temperature model is an effective average over the Brillouin zone. 
However, our results clearly support the view that the contributions to this average are strongly $(\vec{q},\omega)$-dependent. 
In particular, the relaxation observed in Fig.~\ref{fig:relaxation_intensities_4ps_sparse}(b) (also Fig.~\ref{fig:relaxation_intensities_scans_3d}(b, c)) mostly occurs within the first picosecond, whereas the phonons excited at the X high-symmetry point exhibit a much longer relaxation period. 
This observation aligns with the time scales reported by Pankratova~\etal{}~\cite{Pankratova_heat-conserving_2022} as well as with the mode-dependent relaxation times due to phonon-phonon interactions reported by Ritzmann~\etal{}~\cite{Ritzmann_theory_2020}.

\section{Summary and conclusions}

In this work, we presented an approach for exploring phonon relaxation processes through molecular dynamics simulations combined with frozen-phonon multislice calculations. 
This method enables a $(\vec{q}, \omega)$-resolved analysis of \highlight{explicit} phonon dynamics, emphasizing the mode-dependence of relaxation times. 
By introducing the \textit{frozen-trajectory excitation} technique, we achieved efficient and flexible manipulation of atomic trajectories, allowing selective excitation of specific phonon modes.

Our findings for face-centered cubic nickel demonstrate that the relaxation process is highly dependent on the phonon mode. 
In particular, excitations at the X-point exhibit slower relaxation compared to intermediate excitations along the $\Delta$ high-symmetry line. 
This mode-specific dependence highlights that the approaches to phonon relaxation that average over the entire Brillouin zone may obscure important mode-resolved behaviors in real materials.
We see expected oscillatory behavior observable in simulated time-resolved transmission electron microscopy\,\textemdash{}\,related to sample count, excitation region, and integration aperture.
These results offer new avenues for investigating ultrafast phonon dynamics in real materials and will contribute to a deeper understanding of the role of phonons in ultrafast dynamics and energy dissipation mechanisms.
Future extensions could focus on the incorporation of the spin system via consideration of explicit spin-lattice dynamics using methods of Ref.~\cite{hellsvik_general_2019} or heat-conserving three-temperature model calculations \cite{Pankratova_heat-conserving_2022}, by including the effects of electron scattering on magnetic moments \cite{edstrom_elastic_2016,Castellanos-Reyes_unveiling_2023}.

\moved{\highlight{Computationally, by combining GPU acceleration with FTE introduced in this article, we achieve} an efficiency improvement of three to four orders of magnitude over traditional CPU-only calculations, which rely on a one-excitation-per-trajectory approach.}

\begin{acknowledgments}

W. M. acknowledges financial support from the Polish National Agency for Academic Exchange under decision BPN/BEK/2022/1/00179/DEC/1. 
We acknowledge support from the Swedish Research Council, the Olle Engkvist Foundation, and the Knut and Alice Wallenberg Foundation. 
The simulations were enabled by resources provided by the National Academic Infrastructure for Supercomputing in Sweden (NAISS) at NSC Centre, partially funded by the Swedish Research Council through grant agreement no. 2022-06725.
We are grateful to Paul Zeiger for his valuable comments on the manuscript and to Peter M. Oppeneer for his insights and stimulating discussions.

\end{acknowledgments}

\highlight{\section*{Author Contributions statement}}

\highlight{\textbf{Wojciech Marciniak}: conceptualization, formal analysis, data curation, funding acquisition, investigation, methodology, software, visualization, and writing\,\textemdash\,original draft.
\textbf{Joanna Marciniak}: data curation, investigation, validation, and visualization.
\textbf{Jos\'e \'Angel Castellanos-Reyes}: formal analysis, and metodology.
\textbf{J\'an Rusz}: conceptualization, funding acquisition, project administration, resources, and supervision.
\textbf{All authors} contributed to the review and editing.}

\appendix

\highlight{\section{\label{app:details}Detailed description of calculations procedure}}


\highlight{This Appendix outlines the implementation details of the computational protocol described in the main text. Figure~\ref{fig:schematic-full}, an expanded version of Fig.~\ref{fig:schematic}, provides a schematic overview of the workflow, illustrating the data flow, coordinate transformations, and simulation steps. The following subsections present the technical details of the implementation.}

\begin{figure}[ht!]
\center
\resizebox{0.8\columnwidth}{!}{%
    \fmoved{\begin{tikzpicture}[rounded/.style={rectangle, draw=black, rounded corners, align=center},empty/.style={align=center}]
    \node[empty,rotate=90] at (-2.5,-2.1) (label1) {Equilibrium trajectory\\ generation};
    \node[empty,rotate=90] at (3.5,-2.1) (label2) {Any basis (conventional fcc)};
    
    \node[rounded, fill=white] at (0,0) (node1) {Equilibrium positions\\ $\{\vec{r}_{0_{i}}\}_{i=1}^{N_{\rm atom}}$};
    \node[rounded, fill=white] at (0,-2) (node2) {Initial guess\\ $\{\vec{r}_{0_{i}}, \vec{v}_i\}_{i=1}^{N_{\rm atom}}$};
    \node[rounded, fill=white] at (0,-4.2) (node3) {Starting point of equilibrium\\ trajectory (@300K)\\ $\{\vec{r}_{i},\vec{v}_i\}_{i=1}^{N_{\rm atom}}$};
    
    \draw[->] (node1) -- (node2) node[empty,midway,right] {Maxwell-Boltzmann\\ velocity assignment};
    \draw[->] (node2) -- (node3) node[empty,midway,right] {NpT + NVT\\ equilibration};

    \begin{scope}[on background layer]
        \node[rounded,fit=(label1)(label2)(node1)(node2)(node3), fill=red!10] (step1) {};
    \end{scope}

    \node[empty,rotate=90] at (-2.7,-10.7) (label3) {Frozen trajectory excitation (FTE)};
    \node[empty,rotate=90] at (3.3,-7.2) (label4) {Basis:\\ primitive (fcc)};
    \node[empty,rotate=90] at (3.3,-10.7) (label5) {Basis:\\ inverse (bcc)};
    \node[empty,rotate=90] at (3.3,-14.2) (label6) {Basis:\\ primitive (fcc)};
    
    \node[rounded, fill=white] at (0,-6.7) (node4) {Displacements in primitive\\ supercell $\{\vec{u}(\vec{a}_i,t)\}_{i=1}^{N_{\rm atom}}$};
    \node[rounded, fill=white] at (0,-8.7) (node5) {Displacements in\\ ($\vec{q},\omega$)-space $\{\vec{u}(\vec{q}_i,\omega)\}_{i=1}^{N_{\rm atom}}$};
    \node[rounded, fill=white] at (0,-10.7) (node6) {Displacements of the\\ excited system $\{\vec{u}\,'(\vec{q}_i,\omega)\}_{i=1}^{N_{\rm atom}}$};
    \node[rounded, fill=white] at (0,-12.7) (node7) {Full trajectory\\ $\{\vec{u}\,'(\vec{q}_i,\omega),\vec{v}\,'(\vec{q}_i,\omega)\}_{i=1}^{N_{\rm atom}}$};
    \node[rounded, fill=white] at (0,-14.7) (node8) {Full trajectory\\ $\{\vec{r_i}'(t),\vec{v_i}'(t)\}_{i=1}^{N_{\rm atom}}$};

    \draw[->] (node4) -- (node5) node[empty,midway,right] {FFT\\ (Change of basis)};
    \draw[->] (node5) -- (node6) node[empty,midway,right] {Filter application};
    \draw[->] (node6) -- (node7) node[empty,midway,right] {Full trajectory\\ reconstruction};
    \draw[->] (node7) -- (node8) node[empty,midway,right] {IFFT\\ (Change of basis)};

    \begin{scope}[on background layer]
        \node[rounded,fit=(label3)(label4)(label5)(label6)(node4)(node5)(node6)(node7)(node8), fill=green!10] (step2) {};
    \end{scope}

    \node[empty,rotate=90] at (-2.7,-18.2) (label7) {Free relaxation};
    \node[empty,rotate=90] at (3.3,-18.2) (label8) {Basis: any\\ (conventional fcc)};

    \node[rounded, fill=white] at (0,-17.2) (node9) {Starting point \\ $\{\vec{r}_i(0),\vec{v}_i(0)\}_{i=1}^{N_{\rm atom}}$};
    \node[rounded, fill=white] at (0,-19.2) (node10) {Configurations\\ $\vec{R}(t) = \{\vec{r}_i(t)\}_{i=1}^{N_{\rm atom}}$};

    \draw[->] (node9) -- (node10) node[empty,midway,right] {Free\\ NVE MD};

    \begin{scope}[on background layer]
        \node[rounded,fit=(label7)(label8)(node9)(node10), fill=blue!10] (step3) {};
    \end{scope}

    \draw[->] (step1) -- (step2) node[empty,midway,left] {save $\{\vec{r}_i(t)\}_{i=1}^{N_{\rm atom}}$\\ from NVT @ 300K} node[empty,midway,right] {Remap atom indices\\ to primitive basis};
    \draw[->] (step2) -- (step3) node[empty,midway,right] {load selected point\\ to \LAMMPS{} (repeat $\times N_S$)} node[empty,midway,left] {Remap atom indices\\ back to \LAMMPS{}};
    \end{tikzpicture}}
}
\caption{\label{fig:schematic-full} 
\highlight{Detailed schematic of the calculations workflow presents steps performed to obtain atomic configurations at different time delays during the relaxation process.
It shows explicitly performed changes of basis from cartesian space $(\vec{r},t)$, via the primitive cells to reciprocal space and frequency domain $(\vec{q},\omega)$, and back.}
}
\end{figure}

\highlight{\subsection{Details of equilibrium trajectory generation}}

\highlight{Beyond what was described in Sec.~\ref{sec:calculations-details}, there are two additional aspects to be taken into consideration.}


\moved{\highlight{First,} to expedite equilibration, we used the overestimated Maxwell-Boltzmann distribution at 600 K.}
%
%
\moved{\highlight{Second, we need sufficiently sampled, continuous, periodic data to avoid spectral leakage and aliasing.
Thanks to the periodic boundary conditions in the simulation, atomic displacements are periodic with respect to $h$, $k$, and $l$. 
Sampling frequency, initial trajectory length, and supercell sizes are set so that the atoms movement is sufficiently sampled according to the needs dictated by desired angular and energy resolution.
The final concern is the periodicity in time.}
Prior to performing FTE, we apply the Tukey window (cosine-tapered window) $W[n]$ as a pre-processing step to reduce the effects of discontinuities at the first/last snapshot~\cite{bloomfield_fourier_2004}.}
\moved{\highlight{For snapshot index $n \in [0, N]$,} we define Tukey window in the explicit form:}
\mmoved{\begin{equation}
\label{eqn:window_tukey}
    W[n] = \begin{cases}
        \frac{1}{2}\left[ 1 - \cos(\frac{2 \pi n}{\alpha N}) \right], & \text{if } 0 \leq n \leq \frac{\alpha N}{2} \\
        1, & \text{if } \frac{\alpha N}{2} \leq n \leq \frac{N}{2} \\
        W[N-n], & \text{otherwise,}
    \end{cases}
\end{equation}}%
\moved{\highlight{where we apply a cosine decay} to the $\alpha = 0.4$ portion of the trajectory both in the beginning and at the end of the equilibrium MD simulation, resulting in the removal of the first and last 99 snapshots from the total of 500 snapshots recorded in this work.}

\highlight{\subsection{Frozen trajectory excitation details}}


\highlight{In the preparatory phase of the FTE protocol, we map the system into a supercell of primitive fcc unit cells, given by the vectors:}
\mmoved{\begin{align*}
\label{eqn:primitive_vectors}
    \vec{a}_1 &= (0.5a, 0.5a, 0) \\
    \vec{a}_2 &= (0.5a, 0, 0.5a) \\
    \vec{a}_3 &= (0, 0.5a, 0.5a).
\end{align*}}

\moved{This way, we obtain 56 $\times$ 56 $\times$ 28 primitive unit cells.}
\highlight{Converting into the primitive basis allows for unfolding the Brillouin zone while handling the phonon band structure during the intermediate FTE steps.}


\moved{\highlight{This way, we obtain the array indexed by} \highlight{$u[\{x=0,y=1,z=2\}][i][h][k][l][n]$\,\textemdash{}\,where $n$ is the snapshot index for time $t$ (MD frame number)}, that forms a data set for 4D $\vec{u}(\vec{r}^{\,0}, t)\rightarrow\vec{u}(\vec{q},\omega)$ and 1D $\vec{u}(t)\rightarrow\vec{u}(\omega)$ Fourier transforms, without requiring additional memory padding.}

\moved{\highlight{From there, we convert to the $\vec{q}, \omega$-space and obtain 4-dimensional Fourier transform amplitudes related to the dispersion as stated in~Eq.~\ref{eqn:phonon_dispersion_equation}. In discrete form, the intensities are obtained as:}
\begin{equation}
\label{eqn:phonon_dispersion_equation_discrete}
    \begin{split}
        I[q_x][q_y][q_z][\omega] = \lvert \vec{u}[q_x][q_y][q_z][\omega] \rvert =\\= \lvert FFT(\vec{u}[h][k][l][n]) \rvert.
    \end{split}
\end{equation}}


\highlight{In the FTE procedure, we use a filter function to artificially enhance phonon populations in a selected region of $(\vec{q},\omega)$-space by a factor of $A$:}
\mhighlight{\begin{equation}
\label{eqn:enhancement}
    \vec{u}' = (1 + (A - 1)F) \cdot \vec{u},
\end{equation}}%
\highlight{where $F \in \left[ 0, 1 \right]$, so that $\lvert u' \rvert \in \left[ \lvert u \rvert, A \lvert u \rvert \right]$}.

\highlight{This band-pass filter $F$ is multiplicative and based on the Tukey window. In our current implementation, the 4D band-pass filter is expressed as a product of four 1D band-pass filters:}
\mmoved{\begin{equation}
\label{eqn:filter_tukey}
    F[h][k][l][n] = F[h] \cdot F[k] \cdot F[l] \cdot F[n],
\end{equation}}
\moved{where for every index $i = \{h, k, l, n\}$ varying between 0 and $N_i$, the partial functions}
\mmoved{\begin{equation}
\label{eqn:filter_tukey_partials}
    F[i] = \begin{cases}
        0\text{, if } \lvert i_0 - \frac{i}{N_i} \rvert > \Delta_i + \alpha_i \\
        1\text{, if } \lvert i_0 - \frac{i}{N_i} \rvert < \Delta_i \\
        \frac{1}{2} \left[ 1 - \cos \left( \frac{\pi (\alpha_i - \lvert i_0 - \frac{i}{N_i} \rvert + \Delta_i)}{\alpha_i} \right) \right]\text{,} \\
        \text{\hspace{4cm}otherwise,} 
    \end{cases}
\end{equation}}%
\moved{specify a window centered at $i_0$, with a width of $\Delta_i$, and decaying over $\alpha_i$, with all three parameters provided as a fraction of the $N_i$ range.}

\moved{When building the filter, we take care to ensure its hermiticity, i.e., to keep inversion symmetry in $(\vec{q}, \omega)$\,\textemdash{}\,to guarantee real-to-complex ($\vec{r},t$)$\rightarrow$($\vec{q},\omega$) FFT and complex-to-real ($\vec{q},\omega$)$\rightarrow$($\vec{r},t$) IFFT.}

\highlight{In this work we have chosen $\Delta_{\vec{q}} = \Delta_t = \alpha_t$ equal to 1\%, and $\alpha_{\vec{q}}$ equal to 5\% of the transform size (see Eq.~(\ref{eqn:enhancement})).}


\moved{\highlight{Up to this point, memory usage is minimized by storing only the atomic positions. 
This enables all calculations to be performed efficiently and scalably using volatile memory (RAM) only.
Prior to the post-processing stage, we obtain a full trajectory with both displacements and velocities for each atom according to Eq.~\ref{eqn:x_to_v_transformation}.
In practice, we employ a swap-and-multiply transform on the real and imaginary FFT components}: 
\begin{equation}
\label{eqn:x_to_v_transformation_implementation}
    \begin{split}
        \Re[\vec{v}(\vec{q}, \omega)] &= \omega \Im[\vec{u}(\vec{q}, \omega)] \\
        \Im[\vec{v}(\vec{q}, \omega)] &= -\omega \Re[\vec{u}(\vec{q}, \omega)].
    \end{split}
\end{equation}
Tests we conducted, where we compared velocities in reconstructed non-excited trajectory to original \LAMMPS{} data, showed less that 0.2\% relative error for such procedure.}

\highlight{\subsection{Free relaxation and data proccessing}}


\highlight{Finally, we calculate electron diffraction patterns.
To present TDS intensity changes in time, we integrate over an aperture described in~Eq.~\ref{eqn:integration}, }
\moved{where the aperture $\Omega$ is an elliptical region and is defined by:
\begin{equation}
\label{eqn:aperture}
    \begin{split}
        \Omega(\vec{q}_\perp): \frac{[(q_x - q_{x0})\cos{\alpha}+(q_y - q_{y0})\sin{\alpha}]^2}{a^2} + \\
        + \frac{[(q_x - q_{x0})\sin{\alpha}-(q_y - q_{y0})\cos{\alpha}]^2}{b^2} \leq 1,
    \end{split}
\end{equation}
Here, $\alpha$ is the tilt angle (135°), and the semiaxes are $a = 4$ and $b = 3$ data pixels ($\sim 0.08$ and $\sim0.06 \text{\,\AA}^{-1}$, respectively). 
\highlight{The $(q_{x0}, q_{y0})$ refers to the central point of the numerical aperture.}}

\moved{Finally, it is noteworthy that, similar to the FRFPMS method, the proposed FTE approach relies on molecular dynamics (MD) simulations \highlight{and multislice calculations, both of} which scale linearly, i.e., $\mathcal{O}(n)$, with the number of atoms in the system.
The \highlight{following} FFT \highlight{trajectory manipulation}, scaling as $\mathcal{O}(n\log{n})$ is, in realistic applications, \highlight{computationally} negligible due to a much lower scaling factor.}

\bibliography{fcc-Ni-phonons,mybibfile}

\begin{thebibliography}{53}%
\makeatletter
\providecommand \@ifxundefined [1]{%
 \@ifx{#1\undefined}
}%
\providecommand \@ifnum [1]{%
 \ifnum #1\expandafter \@firstoftwo
 \else \expandafter \@secondoftwo
 \fi
}%
\providecommand \@ifx [1]{%
 \ifx #1\expandafter \@firstoftwo
 \else \expandafter \@secondoftwo
 \fi
}%
\providecommand \natexlab [1]{#1}%
\providecommand \enquote  [1]{``#1''}%
\providecommand \bibnamefont  [1]{#1}%
\providecommand \bibfnamefont [1]{#1}%
\providecommand \citenamefont [1]{#1}%
\providecommand \href@noop [0]{\@secondoftwo}%
\providecommand \href [0]{\begingroup \@sanitize@url \@href}%
\providecommand \@href[1]{\@@startlink{#1}\@@href}%
\providecommand \@@href[1]{\endgroup#1\@@endlink}%
\providecommand \@sanitize@url [0]{\catcode `\\12\catcode `\$12\catcode
  `\&12\catcode `\#12\catcode `\^12\catcode `\_12\catcode `\%12\relax}%
\providecommand \@@startlink[1]{}%
\providecommand \@@endlink[0]{}%
\providecommand \url  [0]{\begingroup\@sanitize@url \@url }%
\providecommand \@url [1]{\endgroup\@href {#1}{\urlprefix }}%
\providecommand \urlprefix  [0]{URL }%
\providecommand \Eprint [0]{\href }%
\providecommand \doibase [0]{https://doi.org/}%
\providecommand \selectlanguage [0]{\@gobble}%
\providecommand \bibinfo  [0]{\@secondoftwo}%
\providecommand \bibfield  [0]{\@secondoftwo}%
\providecommand \translation [1]{[#1]}%
\providecommand \BibitemOpen [0]{}%
\providecommand \bibitemStop [0]{}%
\providecommand \bibitemNoStop [0]{.\EOS\space}%
\providecommand \EOS [0]{\spacefactor3000\relax}%
\providecommand \BibitemShut  [1]{\csname bibitem#1\endcsname}%
\let\auto@bib@innerbib\@empty
\bibitem [{\citenamefont {Beaurepaire}\ \emph {et~al.}(1996)\citenamefont
  {Beaurepaire}, \citenamefont {Merle}, \citenamefont {Daunois},\ and\
  \citenamefont {Bigot}}]{Beaurepaire_ultrafast_1996}%
  \BibitemOpen
  \bibfield  {author} {\bibinfo {author} {\bibfnamefont {E.}~\bibnamefont
  {Beaurepaire}}, \bibinfo {author} {\bibfnamefont {J.-C.}\ \bibnamefont
  {Merle}}, \bibinfo {author} {\bibfnamefont {A.}~\bibnamefont {Daunois}},\
  and\ \bibinfo {author} {\bibfnamefont {J.-Y.}\ \bibnamefont {Bigot}},\ }\href
  {https://doi.org/10.1103/PhysRevLett.76.4250} {\bibfield  {journal} {\bibinfo
   {journal} {Phys. Rev. Lett.}\ }\textbf {\bibinfo {volume} {76}},\ \bibinfo
  {pages} {4250} (\bibinfo {year} {1996})}\BibitemShut {NoStop}%
\bibitem [{\citenamefont {Koopmans}\ \emph {et~al.}(2010)\citenamefont
  {Koopmans}, \citenamefont {Malinowski}, \citenamefont {Dalla~Longa},
  \citenamefont {Steiauf}, \citenamefont {F{\"a}hnle}, \citenamefont {Roth},
  \citenamefont {Cinchetti},\ and\ \citenamefont
  {Aeschlimann}}]{Koopmans_explaining_2010}%
  \BibitemOpen
  \bibfield  {author} {\bibinfo {author} {\bibfnamefont {B.}~\bibnamefont
  {Koopmans}}, \bibinfo {author} {\bibfnamefont {G.}~\bibnamefont
  {Malinowski}}, \bibinfo {author} {\bibfnamefont {F.}~\bibnamefont
  {Dalla~Longa}}, \bibinfo {author} {\bibfnamefont {D.}~\bibnamefont
  {Steiauf}}, \bibinfo {author} {\bibfnamefont {M.}~\bibnamefont {F{\"a}hnle}},
  \bibinfo {author} {\bibfnamefont {T.}~\bibnamefont {Roth}}, \bibinfo {author}
  {\bibfnamefont {M.}~\bibnamefont {Cinchetti}},\ and\ \bibinfo {author}
  {\bibfnamefont {M.}~\bibnamefont {Aeschlimann}},\ }\href
  {https://doi.org/10.1038/nmat2593} {\bibfield  {journal} {\bibinfo  {journal}
  {Nat. Mater.}\ }\textbf {\bibinfo {volume} {9}},\ \bibinfo {pages} {259}
  (\bibinfo {year} {2010})}\BibitemShut {NoStop}%
\bibitem [{\citenamefont {Zahn}\ \emph {et~al.}(2021)\citenamefont {Zahn},
  \citenamefont {Jakobs}, \citenamefont {Windsor}, \citenamefont {Seiler},
  \citenamefont {Vasileiadis}, \citenamefont {Butcher}, \citenamefont {Qi},
  \citenamefont {Engel}, \citenamefont {Atxitia}, \citenamefont {Vorberger},\
  and\ \citenamefont {Ernstorfer}}]{Zahn_lattice_2021}%
  \BibitemOpen
  \bibfield  {author} {\bibinfo {author} {\bibfnamefont {D.}~\bibnamefont
  {Zahn}}, \bibinfo {author} {\bibfnamefont {F.}~\bibnamefont {Jakobs}},
  \bibinfo {author} {\bibfnamefont {Y.~W.}\ \bibnamefont {Windsor}}, \bibinfo
  {author} {\bibfnamefont {H.}~\bibnamefont {Seiler}}, \bibinfo {author}
  {\bibfnamefont {T.}~\bibnamefont {Vasileiadis}}, \bibinfo {author}
  {\bibfnamefont {T.~A.}\ \bibnamefont {Butcher}}, \bibinfo {author}
  {\bibfnamefont {Y.}~\bibnamefont {Qi}}, \bibinfo {author} {\bibfnamefont
  {D.}~\bibnamefont {Engel}}, \bibinfo {author} {\bibfnamefont
  {U.}~\bibnamefont {Atxitia}}, \bibinfo {author} {\bibfnamefont
  {J.}~\bibnamefont {Vorberger}},\ and\ \bibinfo {author} {\bibfnamefont
  {R.}~\bibnamefont {Ernstorfer}},\ }\href
  {https://doi.org/10.1103/PhysRevResearch.3.023032} {\bibfield  {journal}
  {\bibinfo  {journal} {Phys. Rev. Res.}\ }\textbf {\bibinfo {volume} {3}},\
  \bibinfo {pages} {023032} (\bibinfo {year} {2021})}\BibitemShut {NoStop}%
\bibitem [{\citenamefont {Pankratova}\ \emph {et~al.}(2022)\citenamefont
  {Pankratova}, \citenamefont {Miranda}, \citenamefont {Thonig}, \citenamefont
  {Pereiro}, \citenamefont {Sj\"oqvist}, \citenamefont {Delin}, \citenamefont
  {Eriksson},\ and\ \citenamefont {Bergman}}]{Pankratova_heat-conserving_2022}%
  \BibitemOpen
  \bibfield  {author} {\bibinfo {author} {\bibfnamefont {M.}~\bibnamefont
  {Pankratova}}, \bibinfo {author} {\bibfnamefont {I.~P.}\ \bibnamefont
  {Miranda}}, \bibinfo {author} {\bibfnamefont {D.}~\bibnamefont {Thonig}},
  \bibinfo {author} {\bibfnamefont {M.}~\bibnamefont {Pereiro}}, \bibinfo
  {author} {\bibfnamefont {E.}~\bibnamefont {Sj\"oqvist}}, \bibinfo {author}
  {\bibfnamefont {A.}~\bibnamefont {Delin}}, \bibinfo {author} {\bibfnamefont
  {O.}~\bibnamefont {Eriksson}},\ and\ \bibinfo {author} {\bibfnamefont
  {A.}~\bibnamefont {Bergman}},\ }\href
  {https://doi.org/10.1103/PhysRevB.106.174407} {\bibfield  {journal} {\bibinfo
   {journal} {Phys. Rev. B}\ }\textbf {\bibinfo {volume} {106}},\ \bibinfo
  {pages} {174407} (\bibinfo {year} {2022})}\BibitemShut {NoStop}%
\bibitem [{\citenamefont {Battiato}\ \emph {et~al.}(2010)\citenamefont
  {Battiato}, \citenamefont {Carva},\ and\ \citenamefont
  {Oppeneer}}]{Battiato_superdiffusive_2010}%
  \BibitemOpen
  \bibfield  {author} {\bibinfo {author} {\bibfnamefont {M.}~\bibnamefont
  {Battiato}}, \bibinfo {author} {\bibfnamefont {K.}~\bibnamefont {Carva}},\
  and\ \bibinfo {author} {\bibfnamefont {P.~M.}\ \bibnamefont {Oppeneer}},\
  }\href {https://doi.org/10.1103/PhysRevLett.105.027203} {\bibfield  {journal}
  {\bibinfo  {journal} {Phys. Rev. Lett.}\ }\textbf {\bibinfo {volume} {105}},\
  \bibinfo {pages} {027203} (\bibinfo {year} {2010})}\BibitemShut {NoStop}%
\bibitem [{\citenamefont {Roth}\ \emph {et~al.}(2012)\citenamefont {Roth},
  \citenamefont {Schellekens}, \citenamefont {Alebrand}, \citenamefont
  {Schmitt}, \citenamefont {Steil}, \citenamefont {Koopmans}, \citenamefont
  {Cinchetti},\ and\ \citenamefont {Aeschlimann}}]{Roth_temperature_2012}%
  \BibitemOpen
  \bibfield  {author} {\bibinfo {author} {\bibfnamefont {T.}~\bibnamefont
  {Roth}}, \bibinfo {author} {\bibfnamefont {A.~J.}\ \bibnamefont
  {Schellekens}}, \bibinfo {author} {\bibfnamefont {S.}~\bibnamefont
  {Alebrand}}, \bibinfo {author} {\bibfnamefont {O.}~\bibnamefont {Schmitt}},
  \bibinfo {author} {\bibfnamefont {D.}~\bibnamefont {Steil}}, \bibinfo
  {author} {\bibfnamefont {B.}~\bibnamefont {Koopmans}}, \bibinfo {author}
  {\bibfnamefont {M.}~\bibnamefont {Cinchetti}},\ and\ \bibinfo {author}
  {\bibfnamefont {M.}~\bibnamefont {Aeschlimann}},\ }\href
  {https://doi.org/10.1103/PhysRevX.2.021006} {\bibfield  {journal} {\bibinfo
  {journal} {Phys. Rev. X}\ }\textbf {\bibinfo {volume} {2}},\ \bibinfo {pages}
  {021006} (\bibinfo {year} {2012})}\BibitemShut {NoStop}%
\bibitem [{\citenamefont {Cheng}\ \emph {et~al.}(2024)\citenamefont {Cheng},
  \citenamefont {Zong}, \citenamefont {Wu}, \citenamefont {Meng}, \citenamefont
  {Xia}, \citenamefont {Qi}, \citenamefont {Zhu}, \citenamefont {Zou},
  \citenamefont {Jiang}, \citenamefont {Guo}, \citenamefont {van Wezel},
  \citenamefont {Kogar}, \citenamefont {Zuerch}, \citenamefont {Zhang},
  \citenamefont {Zhu},\ and\ \citenamefont {Xiang}}]{cheng_ultrafast_2024}%
  \BibitemOpen
  \bibfield  {author} {\bibinfo {author} {\bibfnamefont {Y.}~\bibnamefont
  {Cheng}}, \bibinfo {author} {\bibfnamefont {A.}~\bibnamefont {Zong}},
  \bibinfo {author} {\bibfnamefont {L.}~\bibnamefont {Wu}}, \bibinfo {author}
  {\bibfnamefont {Q.}~\bibnamefont {Meng}}, \bibinfo {author} {\bibfnamefont
  {W.}~\bibnamefont {Xia}}, \bibinfo {author} {\bibfnamefont {F.}~\bibnamefont
  {Qi}}, \bibinfo {author} {\bibfnamefont {P.}~\bibnamefont {Zhu}}, \bibinfo
  {author} {\bibfnamefont {X.}~\bibnamefont {Zou}}, \bibinfo {author}
  {\bibfnamefont {T.}~\bibnamefont {Jiang}}, \bibinfo {author} {\bibfnamefont
  {Y.}~\bibnamefont {Guo}}, \bibinfo {author} {\bibfnamefont {J.}~\bibnamefont
  {van Wezel}}, \bibinfo {author} {\bibfnamefont {A.}~\bibnamefont {Kogar}},
  \bibinfo {author} {\bibfnamefont {M.~W.}\ \bibnamefont {Zuerch}}, \bibinfo
  {author} {\bibfnamefont {J.}~\bibnamefont {Zhang}}, \bibinfo {author}
  {\bibfnamefont {Y.}~\bibnamefont {Zhu}},\ and\ \bibinfo {author}
  {\bibfnamefont {D.}~\bibnamefont {Xiang}},\ }\href
  {https://doi.org/10.1038/s41567-023-02279-x} {\bibfield  {journal} {\bibinfo
  {journal} {Nat. Phys.}\ }\textbf {\bibinfo {volume} {20}},\ \bibinfo {pages}
  {54} (\bibinfo {year} {2024})}\BibitemShut {NoStop}%
\bibitem [{\citenamefont {Hui}\ \emph {et~al.}(2024)\citenamefont {Hui},
  \citenamefont {Alqattan}, \citenamefont {Sennary}, \citenamefont {Golubev},\
  and\ \citenamefont {Hassan}}]{hui_attosecond_2024}%
  \BibitemOpen
  \bibfield  {author} {\bibinfo {author} {\bibfnamefont {D.}~\bibnamefont
  {Hui}}, \bibinfo {author} {\bibfnamefont {H.}~\bibnamefont {Alqattan}},
  \bibinfo {author} {\bibfnamefont {M.}~\bibnamefont {Sennary}}, \bibinfo
  {author} {\bibfnamefont {N.~V.}\ \bibnamefont {Golubev}},\ and\ \bibinfo
  {author} {\bibfnamefont {M.~T.}\ \bibnamefont {Hassan}},\ }\href
  {https://doi.org/10.1126/sciadv.adp5805} {\bibfield  {journal} {\bibinfo
  {journal} {Sci. Adv.}\ }\textbf {\bibinfo {volume} {10}},\ \bibinfo {pages}
  {eadp5805} (\bibinfo {year} {2024})}\BibitemShut {NoStop}%
\bibitem [{\citenamefont {Dettori}\ \emph {et~al.}(2017)\citenamefont
  {Dettori}, \citenamefont {Ceriotti}, \citenamefont {Hunger}, \citenamefont
  {Melis}, \citenamefont {Colombo},\ and\ \citenamefont
  {Donadio}}]{dettori_simulating_2017}%
  \BibitemOpen
  \bibfield  {author} {\bibinfo {author} {\bibfnamefont {R.}~\bibnamefont
  {Dettori}}, \bibinfo {author} {\bibfnamefont {M.}~\bibnamefont {Ceriotti}},
  \bibinfo {author} {\bibfnamefont {J.}~\bibnamefont {Hunger}}, \bibinfo
  {author} {\bibfnamefont {C.}~\bibnamefont {Melis}}, \bibinfo {author}
  {\bibfnamefont {L.}~\bibnamefont {Colombo}},\ and\ \bibinfo {author}
  {\bibfnamefont {D.}~\bibnamefont {Donadio}},\ }\href
  {https://doi.org/10.1021/acs.jctc.6b01108} {\bibfield  {journal} {\bibinfo
  {journal} {J. Chem. Theory Comput.}\ }\textbf {\bibinfo {volume} {13}},\
  \bibinfo {pages} {1284} (\bibinfo {year} {2017})},\ \bibinfo {note} {pMID:
  28112932}\BibitemShut {NoStop}%
\bibitem [{\citenamefont {Dornes}\ \emph {et~al.}(2019)\citenamefont {Dornes},
  \citenamefont {Acremann}, \citenamefont {Savoini}, \citenamefont {Kubli},
  \citenamefont {Neugebauer}, \citenamefont {Abreu}, \citenamefont {Huber},
  \citenamefont {Lantz}, \citenamefont {Vaz}, \citenamefont {Lemke},
  \citenamefont {Bothschafter}, \citenamefont {Porer}, \citenamefont
  {Esposito}, \citenamefont {Rettig}, \citenamefont {Buzzi}, \citenamefont
  {Alberca}, \citenamefont {Windsor}, \citenamefont {Beaud}, \citenamefont
  {Staub}, \citenamefont {Zhu}, \citenamefont {Song}, \citenamefont {Glownia},\
  and\ \citenamefont {Johnson}}]{Dornes_einstein-de-haas_2019}%
  \BibitemOpen
  \bibfield  {author} {\bibinfo {author} {\bibfnamefont {C.}~\bibnamefont
  {Dornes}}, \bibinfo {author} {\bibfnamefont {Y.}~\bibnamefont {Acremann}},
  \bibinfo {author} {\bibfnamefont {M.}~\bibnamefont {Savoini}}, \bibinfo
  {author} {\bibfnamefont {M.}~\bibnamefont {Kubli}}, \bibinfo {author}
  {\bibfnamefont {M.~J.}\ \bibnamefont {Neugebauer}}, \bibinfo {author}
  {\bibfnamefont {E.}~\bibnamefont {Abreu}}, \bibinfo {author} {\bibfnamefont
  {L.}~\bibnamefont {Huber}}, \bibinfo {author} {\bibfnamefont
  {G.}~\bibnamefont {Lantz}}, \bibinfo {author} {\bibfnamefont {C.~A.~F.}\
  \bibnamefont {Vaz}}, \bibinfo {author} {\bibfnamefont {H.}~\bibnamefont
  {Lemke}}, \bibinfo {author} {\bibfnamefont {E.~M.}\ \bibnamefont
  {Bothschafter}}, \bibinfo {author} {\bibfnamefont {M.}~\bibnamefont {Porer}},
  \bibinfo {author} {\bibfnamefont {V.}~\bibnamefont {Esposito}}, \bibinfo
  {author} {\bibfnamefont {L.}~\bibnamefont {Rettig}}, \bibinfo {author}
  {\bibfnamefont {M.}~\bibnamefont {Buzzi}}, \bibinfo {author} {\bibfnamefont
  {A.}~\bibnamefont {Alberca}}, \bibinfo {author} {\bibfnamefont {Y.~W.}\
  \bibnamefont {Windsor}}, \bibinfo {author} {\bibfnamefont {P.}~\bibnamefont
  {Beaud}}, \bibinfo {author} {\bibfnamefont {U.}~\bibnamefont {Staub}},
  \bibinfo {author} {\bibfnamefont {D.}~\bibnamefont {Zhu}}, \bibinfo {author}
  {\bibfnamefont {S.}~\bibnamefont {Song}}, \bibinfo {author} {\bibfnamefont
  {J.~M.}\ \bibnamefont {Glownia}},\ and\ \bibinfo {author} {\bibfnamefont
  {S.~L.}\ \bibnamefont {Johnson}},\ }\href
  {https://doi.org/10.1038/s41586-018-0822-7} {\bibfield  {journal} {\bibinfo
  {journal} {Nature}\ }\textbf {\bibinfo {volume} {565}},\ \bibinfo {pages}
  {209} (\bibinfo {year} {2019})}\BibitemShut {NoStop}%
\bibitem [{\citenamefont {Maldonado}\ \emph {et~al.}(2020)\citenamefont
  {Maldonado}, \citenamefont {Chase}, \citenamefont {Reid}, \citenamefont
  {Shen}, \citenamefont {Li}, \citenamefont {Carva}, \citenamefont {Payer},
  \citenamefont {Horn~von Hoegen}, \citenamefont {Sokolowski-Tinten},
  \citenamefont {Wang}, \citenamefont {Oppeneer},\ and\ \citenamefont
  {D\"urr}}]{Maldonado_tracking_2020}%
  \BibitemOpen
  \bibfield  {author} {\bibinfo {author} {\bibfnamefont {P.}~\bibnamefont
  {Maldonado}}, \bibinfo {author} {\bibfnamefont {T.}~\bibnamefont {Chase}},
  \bibinfo {author} {\bibfnamefont {A.~H.}\ \bibnamefont {Reid}}, \bibinfo
  {author} {\bibfnamefont {X.}~\bibnamefont {Shen}}, \bibinfo {author}
  {\bibfnamefont {R.~K.}\ \bibnamefont {Li}}, \bibinfo {author} {\bibfnamefont
  {K.}~\bibnamefont {Carva}}, \bibinfo {author} {\bibfnamefont
  {T.}~\bibnamefont {Payer}}, \bibinfo {author} {\bibfnamefont
  {M.}~\bibnamefont {Horn~von Hoegen}}, \bibinfo {author} {\bibfnamefont
  {K.}~\bibnamefont {Sokolowski-Tinten}}, \bibinfo {author} {\bibfnamefont
  {X.~J.}\ \bibnamefont {Wang}}, \bibinfo {author} {\bibfnamefont {P.~M.}\
  \bibnamefont {Oppeneer}},\ and\ \bibinfo {author} {\bibfnamefont {H.~A.}\
  \bibnamefont {D\"urr}},\ }\href {https://doi.org/10.1103/PhysRevB.101.100302}
  {\bibfield  {journal} {\bibinfo  {journal} {Phys. Rev. B}\ }\textbf {\bibinfo
  {volume} {101}},\ \bibinfo {pages} {100302} (\bibinfo {year}
  {2020})}\BibitemShut {NoStop}%
\bibitem [{\citenamefont {Ritzmann}\ \emph {et~al.}(2020)\citenamefont
  {Ritzmann}, \citenamefont {Oppeneer},\ and\ \citenamefont
  {Maldonado}}]{Ritzmann_theory_2020}%
  \BibitemOpen
  \bibfield  {author} {\bibinfo {author} {\bibfnamefont {U.}~\bibnamefont
  {Ritzmann}}, \bibinfo {author} {\bibfnamefont {P.~M.}\ \bibnamefont
  {Oppeneer}},\ and\ \bibinfo {author} {\bibfnamefont {P.}~\bibnamefont
  {Maldonado}},\ }\href {https://doi.org/10.1103/PhysRevB.102.214305}
  {\bibfield  {journal} {\bibinfo  {journal} {Phys. Rev. B}\ }\textbf {\bibinfo
  {volume} {102}},\ \bibinfo {pages} {214305} (\bibinfo {year}
  {2020})}\BibitemShut {NoStop}%
\bibitem [{\citenamefont {Tauchert}\ \emph {et~al.}(2022)\citenamefont
  {Tauchert}, \citenamefont {Volkov}, \citenamefont {Ehberger}, \citenamefont
  {Kazenwadel}, \citenamefont {Evers}, \citenamefont {Lange}, \citenamefont
  {Donges}, \citenamefont {Book}, \citenamefont {Kreuzpaintner}, \citenamefont
  {Nowak},\ and\ \citenamefont {Baum}}]{Tauchert_polarized_2022}%
  \BibitemOpen
  \bibfield  {author} {\bibinfo {author} {\bibfnamefont {S.~R.}\ \bibnamefont
  {Tauchert}}, \bibinfo {author} {\bibfnamefont {M.}~\bibnamefont {Volkov}},
  \bibinfo {author} {\bibfnamefont {D.}~\bibnamefont {Ehberger}}, \bibinfo
  {author} {\bibfnamefont {D.}~\bibnamefont {Kazenwadel}}, \bibinfo {author}
  {\bibfnamefont {M.}~\bibnamefont {Evers}}, \bibinfo {author} {\bibfnamefont
  {H.}~\bibnamefont {Lange}}, \bibinfo {author} {\bibfnamefont
  {A.}~\bibnamefont {Donges}}, \bibinfo {author} {\bibfnamefont
  {A.}~\bibnamefont {Book}}, \bibinfo {author} {\bibfnamefont {W.}~\bibnamefont
  {Kreuzpaintner}}, \bibinfo {author} {\bibfnamefont {U.}~\bibnamefont
  {Nowak}},\ and\ \bibinfo {author} {\bibfnamefont {P.}~\bibnamefont {Baum}},\
  }\href {https://doi.org/10.1038/s41586-021-04306-4} {\bibfield  {journal}
  {\bibinfo  {journal} {Nature}\ }\textbf {\bibinfo {volume} {602}},\ \bibinfo
  {pages} {73} (\bibinfo {year} {2022})}\BibitemShut {NoStop}%
\bibitem [{\citenamefont {Barantani}\ \emph {et~al.}(2024)\citenamefont
  {Barantani}, \citenamefont {Claude}, \citenamefont {Iyikanat}, \citenamefont
  {Madan}, \citenamefont {Sapozhnik}, \citenamefont {Puppin}, \citenamefont
  {Weaver}, \citenamefont {LaGrange}, \citenamefont {de~Abajo},\ and\
  \citenamefont {Carbone}}]{barantani_ultrafast_2024}%
  \BibitemOpen
  \bibfield  {author} {\bibinfo {author} {\bibfnamefont {F.}~\bibnamefont
  {Barantani}}, \bibinfo {author} {\bibfnamefont {R.}~\bibnamefont {Claude}},
  \bibinfo {author} {\bibfnamefont {F.}~\bibnamefont {Iyikanat}}, \bibinfo
  {author} {\bibfnamefont {I.}~\bibnamefont {Madan}}, \bibinfo {author}
  {\bibfnamefont {A.~A.}\ \bibnamefont {Sapozhnik}}, \bibinfo {author}
  {\bibfnamefont {M.}~\bibnamefont {Puppin}}, \bibinfo {author} {\bibfnamefont
  {B.}~\bibnamefont {Weaver}}, \bibinfo {author} {\bibfnamefont
  {T.}~\bibnamefont {LaGrange}}, \bibinfo {author} {\bibfnamefont {F.~J.~G.}\
  \bibnamefont {de~Abajo}},\ and\ \bibinfo {author} {\bibfnamefont
  {F.}~\bibnamefont {Carbone}},\ }\href@noop {} {\bibinfo {title} {Ultrafast
  momentum-resolved visualization of the interplay between phonon-mediated
  scattering and plasmons in graphite}} (\bibinfo {year} {2024}),\ \Eprint
  {https://arxiv.org/abs/2410.06810} {arXiv:2410.06810} \BibitemShut {NoStop}%
\bibitem [{\citenamefont {Haider}\ \emph {et~al.}(1998)\citenamefont {Haider},
  \citenamefont {Uhlemann}, \citenamefont {Schwan}, \citenamefont {Rose},
  \citenamefont {Kabius},\ and\ \citenamefont {Urban}}]{Haider_electron_1998}%
  \BibitemOpen
  \bibfield  {author} {\bibinfo {author} {\bibfnamefont {M.}~\bibnamefont
  {Haider}}, \bibinfo {author} {\bibfnamefont {S.}~\bibnamefont {Uhlemann}},
  \bibinfo {author} {\bibfnamefont {E.}~\bibnamefont {Schwan}}, \bibinfo
  {author} {\bibfnamefont {H.}~\bibnamefont {Rose}}, \bibinfo {author}
  {\bibfnamefont {B.}~\bibnamefont {Kabius}},\ and\ \bibinfo {author}
  {\bibfnamefont {K.}~\bibnamefont {Urban}},\ }\href
  {https://doi.org/10.1038/33823} {\bibfield  {journal} {\bibinfo  {journal}
  {Nature}\ }\textbf {\bibinfo {volume} {392}},\ \bibinfo {pages} {768}
  (\bibinfo {year} {1998})}\BibitemShut {NoStop}%
\bibitem [{\citenamefont {Batson}\ \emph {et~al.}(2002)\citenamefont {Batson},
  \citenamefont {Dellby},\ and\ \citenamefont
  {Krivanek}}]{Batson_sub-Angstrom_2002}%
  \BibitemOpen
  \bibfield  {author} {\bibinfo {author} {\bibfnamefont {P.~E.}\ \bibnamefont
  {Batson}}, \bibinfo {author} {\bibfnamefont {N.}~\bibnamefont {Dellby}},\
  and\ \bibinfo {author} {\bibfnamefont {O.~L.}\ \bibnamefont {Krivanek}},\
  }\href {https://doi.org/10.1038/nature00972} {\bibfield  {journal} {\bibinfo
  {journal} {Nature}\ }\textbf {\bibinfo {volume} {418}},\ \bibinfo {pages}
  {617} (\bibinfo {year} {2002})}\BibitemShut {NoStop}%
\bibitem [{\citenamefont {Krivanek}\ \emph {et~al.}(2014)\citenamefont
  {Krivanek}, \citenamefont {Lovejoy}, \citenamefont {Dellby}, \citenamefont
  {Aoki}, \citenamefont {Carpenter}, \citenamefont {Rez}, \citenamefont
  {Soignard}, \citenamefont {Zhu}, \citenamefont {Batson}, \citenamefont
  {Lagos}, \citenamefont {Egerton},\ and\ \citenamefont
  {Crozier}}]{Krivanek_vibrational_2014}%
  \BibitemOpen
  \bibfield  {author} {\bibinfo {author} {\bibfnamefont {O.~L.}\ \bibnamefont
  {Krivanek}}, \bibinfo {author} {\bibfnamefont {T.~C.}\ \bibnamefont
  {Lovejoy}}, \bibinfo {author} {\bibfnamefont {N.}~\bibnamefont {Dellby}},
  \bibinfo {author} {\bibfnamefont {T.}~\bibnamefont {Aoki}}, \bibinfo {author}
  {\bibfnamefont {R.~W.}\ \bibnamefont {Carpenter}}, \bibinfo {author}
  {\bibfnamefont {P.}~\bibnamefont {Rez}}, \bibinfo {author} {\bibfnamefont
  {E.}~\bibnamefont {Soignard}}, \bibinfo {author} {\bibfnamefont
  {J.}~\bibnamefont {Zhu}}, \bibinfo {author} {\bibfnamefont {P.~E.}\
  \bibnamefont {Batson}}, \bibinfo {author} {\bibfnamefont {M.~J.}\
  \bibnamefont {Lagos}}, \bibinfo {author} {\bibfnamefont {R.~F.}\ \bibnamefont
  {Egerton}},\ and\ \bibinfo {author} {\bibfnamefont {P.~A.}\ \bibnamefont
  {Crozier}},\ }\href {https://doi.org/10.1038/nature13870} {\bibfield
  {journal} {\bibinfo  {journal} {Nature}\ }\textbf {\bibinfo {volume} {514}},\
  \bibinfo {pages} {209} (\bibinfo {year} {2014})}\BibitemShut {NoStop}%
\bibitem [{\citenamefont {Li}\ \emph {et~al.}(2013)\citenamefont {Li},
  \citenamefont {Mooney}, \citenamefont {Zheng}, \citenamefont {Booth},
  \citenamefont {Braunfeld}, \citenamefont {Gubbens}, \citenamefont {Agard},\
  and\ \citenamefont {Cheng}}]{Li_electron-counting_2013}%
  \BibitemOpen
  \bibfield  {author} {\bibinfo {author} {\bibfnamefont {X.}~\bibnamefont
  {Li}}, \bibinfo {author} {\bibfnamefont {P.}~\bibnamefont {Mooney}}, \bibinfo
  {author} {\bibfnamefont {S.}~\bibnamefont {Zheng}}, \bibinfo {author}
  {\bibfnamefont {C.~R.}\ \bibnamefont {Booth}}, \bibinfo {author}
  {\bibfnamefont {M.~B.}\ \bibnamefont {Braunfeld}}, \bibinfo {author}
  {\bibfnamefont {S.}~\bibnamefont {Gubbens}}, \bibinfo {author} {\bibfnamefont
  {D.~A.}\ \bibnamefont {Agard}},\ and\ \bibinfo {author} {\bibfnamefont
  {Y.}~\bibnamefont {Cheng}},\ }\href {https://doi.org/10.1038/nmeth.2472}
  {\bibfield  {journal} {\bibinfo  {journal} {Nat. Methods}\ }\textbf {\bibinfo
  {volume} {10}},\ \bibinfo {pages} {584} (\bibinfo {year} {2013})}\BibitemShut
  {NoStop}%
\bibitem [{\citenamefont {Hart}\ \emph {et~al.}(2017)\citenamefont {Hart},
  \citenamefont {Lang}, \citenamefont {Leff}, \citenamefont {Longo},
  \citenamefont {Trevor}, \citenamefont {Twesten},\ and\ \citenamefont
  {Taheri}}]{Hart_direct_2017}%
  \BibitemOpen
  \bibfield  {author} {\bibinfo {author} {\bibfnamefont {J.~L.}\ \bibnamefont
  {Hart}}, \bibinfo {author} {\bibfnamefont {A.~C.}\ \bibnamefont {Lang}},
  \bibinfo {author} {\bibfnamefont {A.~C.}\ \bibnamefont {Leff}}, \bibinfo
  {author} {\bibfnamefont {P.}~\bibnamefont {Longo}}, \bibinfo {author}
  {\bibfnamefont {C.}~\bibnamefont {Trevor}}, \bibinfo {author} {\bibfnamefont
  {R.~D.}\ \bibnamefont {Twesten}},\ and\ \bibinfo {author} {\bibfnamefont
  {M.~L.}\ \bibnamefont {Taheri}},\ }\href
  {https://doi.org/10.1038/s41598-017-07709-4} {\bibfield  {journal} {\bibinfo
  {journal} {Sci. Rep.}\ }\textbf {\bibinfo {volume} {7}},\ \bibinfo {pages}
  {8243} (\bibinfo {year} {2017})}\BibitemShut {NoStop}%
\bibitem [{\citenamefont {Zewail}(2010)}]{Zewail_four-dimensional_2010}%
  \BibitemOpen
  \bibfield  {author} {\bibinfo {author} {\bibfnamefont {A.~H.}\ \bibnamefont
  {Zewail}},\ }\href {https://doi.org/10.1126/science.1166135} {\bibfield
  {journal} {\bibinfo  {journal} {Science}\ }\textbf {\bibinfo {volume}
  {328}},\ \bibinfo {pages} {187} (\bibinfo {year} {2010})}\BibitemShut
  {NoStop}%
\bibitem [{\citenamefont {Feist}\ \emph {et~al.}(2017)\citenamefont {Feist},
  \citenamefont {Bach}, \citenamefont {{Rubiano da Silva}}, \citenamefont
  {Danz}, \citenamefont {Möller}, \citenamefont {Priebe}, \citenamefont
  {Domröse}, \citenamefont {Gatzmann}, \citenamefont {Rost}, \citenamefont
  {Schauss}, \citenamefont {Strauch}, \citenamefont {Bormann}, \citenamefont
  {Sivis}, \citenamefont {Schäfer},\ and\ \citenamefont
  {Ropers}}]{Feist_ultrafast_2017}%
  \BibitemOpen
  \bibfield  {author} {\bibinfo {author} {\bibfnamefont {A.}~\bibnamefont
  {Feist}}, \bibinfo {author} {\bibfnamefont {N.}~\bibnamefont {Bach}},
  \bibinfo {author} {\bibfnamefont {N.}~\bibnamefont {{Rubiano da Silva}}},
  \bibinfo {author} {\bibfnamefont {T.}~\bibnamefont {Danz}}, \bibinfo {author}
  {\bibfnamefont {M.}~\bibnamefont {Möller}}, \bibinfo {author} {\bibfnamefont
  {K.~E.}\ \bibnamefont {Priebe}}, \bibinfo {author} {\bibfnamefont
  {T.}~\bibnamefont {Domröse}}, \bibinfo {author} {\bibfnamefont {J.~G.}\
  \bibnamefont {Gatzmann}}, \bibinfo {author} {\bibfnamefont {S.}~\bibnamefont
  {Rost}}, \bibinfo {author} {\bibfnamefont {J.}~\bibnamefont {Schauss}},
  \bibinfo {author} {\bibfnamefont {S.}~\bibnamefont {Strauch}}, \bibinfo
  {author} {\bibfnamefont {R.}~\bibnamefont {Bormann}}, \bibinfo {author}
  {\bibfnamefont {M.}~\bibnamefont {Sivis}}, \bibinfo {author} {\bibfnamefont
  {S.}~\bibnamefont {Schäfer}},\ and\ \bibinfo {author} {\bibfnamefont
  {C.}~\bibnamefont {Ropers}},\ }\href
  {https://doi.org/https://doi.org/10.1016/j.ultramic.2016.12.005} {\bibfield
  {journal} {\bibinfo  {journal} {Ultramicroscopy}\ }\textbf {\bibinfo {volume}
  {176}},\ \bibinfo {pages} {63} (\bibinfo {year} {2017})},\ \bibinfo {note}
  {70th Birthday of Robert Sinclair and 65th Birthday of Nestor J. Zaluzec PICO
  2017 – Fourth Conference on Frontiers of Aberration Corrected Electron
  Microscopy}\BibitemShut {NoStop}%
\bibitem [{\citenamefont {Gaida}\ \emph {et~al.}(2024)\citenamefont {Gaida},
  \citenamefont {Louren{\c{c}}o-Martins}, \citenamefont {Sivis}, \citenamefont
  {Rittmann}, \citenamefont {Feist}, \citenamefont {Garc{\'i}a~de Abajo},\ and\
  \citenamefont {Ropers}}]{Gaida_attosecond_2024}%
  \BibitemOpen
  \bibfield  {author} {\bibinfo {author} {\bibfnamefont {J.~H.}\ \bibnamefont
  {Gaida}}, \bibinfo {author} {\bibfnamefont {H.}~\bibnamefont
  {Louren{\c{c}}o-Martins}}, \bibinfo {author} {\bibfnamefont {M.}~\bibnamefont
  {Sivis}}, \bibinfo {author} {\bibfnamefont {T.}~\bibnamefont {Rittmann}},
  \bibinfo {author} {\bibfnamefont {A.}~\bibnamefont {Feist}}, \bibinfo
  {author} {\bibfnamefont {F.~J.}\ \bibnamefont {Garc{\'i}a~de Abajo}},\ and\
  \bibinfo {author} {\bibfnamefont {C.}~\bibnamefont {Ropers}},\ }\href
  {https://doi.org/10.1038/s41566-024-01380-8} {\bibfield  {journal} {\bibinfo
  {journal} {Nat. Photon.}\ }\textbf {\bibinfo {volume} {18}},\ \bibinfo
  {pages} {509} (\bibinfo {year} {2024})}\BibitemShut {NoStop}%
\bibitem [{\citenamefont {Ye-Jin~Kim}\ and\ \citenamefont
  {Kwon}(2024)}]{Kim_high-resolution_2024}%
  \BibitemOpen
  \bibfield  {author} {\bibinfo {author} {\bibfnamefont {H.-W.~N.}\
  \bibnamefont {Ye-Jin~Kim}, \bibfnamefont {Won-Woo~Park}}\ and\ \bibinfo
  {author} {\bibfnamefont {O.-H.}\ \bibnamefont {Kwon}},\ }\href
  {https://doi.org/10.1080/23746149.2024.2316710} {\bibfield  {journal}
  {\bibinfo  {journal} {Adv. Phys. X}\ }\textbf {\bibinfo {volume} {9}},\
  \bibinfo {pages} {2316710} (\bibinfo {year} {2024})}\BibitemShut {NoStop}%
\bibitem [{\citenamefont {Thompson}\ \emph {et~al.}(2022)\citenamefont
  {Thompson}, \citenamefont {Aktulga}, \citenamefont {Berger}, \citenamefont
  {Bolintineanu}, \citenamefont {Brown}, \citenamefont {Crozier}, \citenamefont
  {{in 't Veld}}, \citenamefont {Kohlmeyer}, \citenamefont {Moore},
  \citenamefont {Nguyen}, \citenamefont {Shan}, \citenamefont {Stevens},
  \citenamefont {Tranchida}, \citenamefont {Trott},\ and\ \citenamefont
  {Plimpton}}]{thompson_lammps_2022}%
  \BibitemOpen
  \bibfield  {author} {\bibinfo {author} {\bibfnamefont {A.~P.}\ \bibnamefont
  {Thompson}}, \bibinfo {author} {\bibfnamefont {H.~M.}\ \bibnamefont
  {Aktulga}}, \bibinfo {author} {\bibfnamefont {R.}~\bibnamefont {Berger}},
  \bibinfo {author} {\bibfnamefont {D.~S.}\ \bibnamefont {Bolintineanu}},
  \bibinfo {author} {\bibfnamefont {W.~M.}\ \bibnamefont {Brown}}, \bibinfo
  {author} {\bibfnamefont {P.~S.}\ \bibnamefont {Crozier}}, \bibinfo {author}
  {\bibfnamefont {P.~J.}\ \bibnamefont {{in 't Veld}}}, \bibinfo {author}
  {\bibfnamefont {A.}~\bibnamefont {Kohlmeyer}}, \bibinfo {author}
  {\bibfnamefont {S.~G.}\ \bibnamefont {Moore}}, \bibinfo {author}
  {\bibfnamefont {T.~D.}\ \bibnamefont {Nguyen}}, \bibinfo {author}
  {\bibfnamefont {R.}~\bibnamefont {Shan}}, \bibinfo {author} {\bibfnamefont
  {M.~J.}\ \bibnamefont {Stevens}}, \bibinfo {author} {\bibfnamefont
  {J.}~\bibnamefont {Tranchida}}, \bibinfo {author} {\bibfnamefont
  {C.}~\bibnamefont {Trott}},\ and\ \bibinfo {author} {\bibfnamefont {S.~J.}\
  \bibnamefont {Plimpton}},\ }\href {https://doi.org/10.1016/j.cpc.2021.108171}
  {\bibfield  {journal} {\bibinfo  {journal} {Comput. Phys. Commun.}\ }\textbf
  {\bibinfo {volume} {271}},\ \bibinfo {pages} {108171} (\bibinfo {year}
  {2022})}\BibitemShut {NoStop}%
\bibitem [{\citenamefont {Eriksson}\ \emph {et~al.}(2017)\citenamefont
  {Eriksson}, \citenamefont {Bergman}, \citenamefont {Bergqvist},\ and\
  \citenamefont {Hellsvik}}]{eriksson_atomistic_2017}%
  \BibitemOpen
  \bibfield  {author} {\bibinfo {author} {\bibfnamefont {O.}~\bibnamefont
  {Eriksson}}, \bibinfo {author} {\bibfnamefont {A.}~\bibnamefont {Bergman}},
  \bibinfo {author} {\bibfnamefont {L.}~\bibnamefont {Bergqvist}},\ and\
  \bibinfo {author} {\bibfnamefont {J.}~\bibnamefont {Hellsvik}},\ }\href
  {https://doi.org/10.1093/oso/9780198788669.001.0001} {\emph {\bibinfo {title}
  {Atomistic {Spin} {Dynamics}: {Foundations} and {Applications}}}}\ (\bibinfo
  {publisher} {Oxford University Press},\ \bibinfo {year} {2017})\BibitemShut
  {NoStop}%
\bibitem [{\citenamefont {Loane}\ \emph {et~al.}(1991)\citenamefont {Loane},
  \citenamefont {Xu},\ and\ \citenamefont {Silcox}}]{loane_thermal_1991}%
  \BibitemOpen
  \bibfield  {author} {\bibinfo {author} {\bibfnamefont {R.~F.}\ \bibnamefont
  {Loane}}, \bibinfo {author} {\bibfnamefont {P.}~\bibnamefont {Xu}},\ and\
  \bibinfo {author} {\bibfnamefont {J.}~\bibnamefont {Silcox}},\ }\href
  {https://doi.org/10.1107/S0108767391000375} {\bibfield  {journal} {\bibinfo
  {journal} {Acta Cryst A}\ }\textbf {\bibinfo {volume} {47}},\ \bibinfo
  {pages} {267} (\bibinfo {year} {1991})}\BibitemShut {NoStop}%
\bibitem [{\citenamefont {Barthel}(2018)}]{barthel_dr_2018}%
  \BibitemOpen
  \bibfield  {author} {\bibinfo {author} {\bibfnamefont {J.}~\bibnamefont
  {Barthel}},\ }\href {https://doi.org/10.1016/j.ultramic.2018.06.003}
  {\bibfield  {journal} {\bibinfo  {journal} {Ultramicroscopy}\ }\textbf
  {\bibinfo {volume} {193}},\ \bibinfo {pages} {1} (\bibinfo {year}
  {2018})}\BibitemShut {NoStop}%
\bibitem [{\citenamefont {Zeiger}\ and\ \citenamefont
  {Rusz}(2020)}]{zeiger_efficient_2020}%
  \BibitemOpen
  \bibfield  {author} {\bibinfo {author} {\bibfnamefont {P.~M.}\ \bibnamefont
  {Zeiger}}\ and\ \bibinfo {author} {\bibfnamefont {J.}~\bibnamefont {Rusz}},\
  }\href {https://doi.org/10.1103/PhysRevLett.124.025501} {\bibfield  {journal}
  {\bibinfo  {journal} {Phys. Rev. Lett.}\ }\textbf {\bibinfo {volume} {124}},\
  \bibinfo {pages} {025501} (\bibinfo {year} {2020})}\BibitemShut {NoStop}%
\bibitem [{\citenamefont {Zeiger}\ and\ \citenamefont
  {Rusz}(2021)}]{zeiger_frequencyresolved_2021}%
  \BibitemOpen
  \bibfield  {author} {\bibinfo {author} {\bibfnamefont {P.~M.}\ \bibnamefont
  {Zeiger}}\ and\ \bibinfo {author} {\bibfnamefont {J.}~\bibnamefont {Rusz}},\
  }\href {https://doi.org/10.1103/PhysRevB.104.104301} {\bibfield  {journal}
  {\bibinfo  {journal} {Phys. Rev. B}\ }\textbf {\bibinfo {volume} {104}},\
  \bibinfo {pages} {104301} (\bibinfo {year} {2021})}\BibitemShut {NoStop}%
\bibitem [{\citenamefont {Frigo}\ and\ \citenamefont
  {Johnson}(2005)}]{frigo_design_2005}%
  \BibitemOpen
  \bibfield  {author} {\bibinfo {author} {\bibfnamefont {M.}~\bibnamefont
  {Frigo}}\ and\ \bibinfo {author} {\bibfnamefont {S.}~\bibnamefont
  {Johnson}},\ }\href {https://doi.org/10.1109/JPROC.2004.840301} {\bibfield
  {journal} {\bibinfo  {journal} {Proc. IEEE}\ }\textbf {\bibinfo {volume}
  {93}},\ \bibinfo {pages} {216} (\bibinfo {year} {2005})}\BibitemShut
  {NoStop}%
\bibitem [{\citenamefont {Forbes}\ \emph {et~al.}(2010)\citenamefont {Forbes},
  \citenamefont {Martin}, \citenamefont {Findlay}, \citenamefont {D'Alfonso},\
  and\ \citenamefont {Allen}}]{forbes_quantum_2010}%
  \BibitemOpen
  \bibfield  {author} {\bibinfo {author} {\bibfnamefont {B.~D.}\ \bibnamefont
  {Forbes}}, \bibinfo {author} {\bibfnamefont {A.~V.}\ \bibnamefont {Martin}},
  \bibinfo {author} {\bibfnamefont {S.~D.}\ \bibnamefont {Findlay}}, \bibinfo
  {author} {\bibfnamefont {A.~J.}\ \bibnamefont {D'Alfonso}},\ and\ \bibinfo
  {author} {\bibfnamefont {L.~J.}\ \bibnamefont {Allen}},\ }\href
  {https://doi.org/10.1103/PhysRevB.82.104103} {\bibfield  {journal} {\bibinfo
  {journal} {Phys. Rev. B}\ }\textbf {\bibinfo {volume} {82}},\ \bibinfo
  {pages} {104103} (\bibinfo {year} {2010})}\BibitemShut {NoStop}%
\bibitem [{\citenamefont {Lugg}\ \emph {et~al.}(2015)\citenamefont {Lugg},
  \citenamefont {Forbes}, \citenamefont {Findlay},\ and\ \citenamefont
  {Allen}}]{lugg_atomic_2015}%
  \BibitemOpen
  \bibfield  {author} {\bibinfo {author} {\bibfnamefont {N.~R.}\ \bibnamefont
  {Lugg}}, \bibinfo {author} {\bibfnamefont {B.~D.}\ \bibnamefont {Forbes}},
  \bibinfo {author} {\bibfnamefont {S.~D.}\ \bibnamefont {Findlay}},\ and\
  \bibinfo {author} {\bibfnamefont {L.~J.}\ \bibnamefont {Allen}},\ }\href
  {https://doi.org/10.1103/PhysRevB.91.144108} {\bibfield  {journal} {\bibinfo
  {journal} {Phys. Rev. B}\ }\textbf {\bibinfo {volume} {91}},\ \bibinfo
  {pages} {144108} (\bibinfo {year} {2015})}\BibitemShut {NoStop}%
\bibitem [{\citenamefont {Löfgren}\ \emph {et~al.}(2016)\citenamefont
  {Löfgren}, \citenamefont {Zeiger}, \citenamefont {Kocevski},\ and\
  \citenamefont {Rusz}}]{lofgren_influence_2016}%
  \BibitemOpen
  \bibfield  {author} {\bibinfo {author} {\bibfnamefont {A.}~\bibnamefont
  {Löfgren}}, \bibinfo {author} {\bibfnamefont {P.}~\bibnamefont {Zeiger}},
  \bibinfo {author} {\bibfnamefont {V.}~\bibnamefont {Kocevski}},\ and\
  \bibinfo {author} {\bibfnamefont {J.}~\bibnamefont {Rusz}},\ }\href
  {https://doi.org/10.1016/j.ultramic.2016.01.007} {\bibfield  {journal}
  {\bibinfo  {journal} {Ultramicroscopy}\ }\textbf {\bibinfo {volume} {164}},\
  \bibinfo {pages} {62} (\bibinfo {year} {2016})}\BibitemShut {NoStop}%
\bibitem [{\citenamefont {Aveyard}\ \emph {et~al.}(2014)\citenamefont
  {Aveyard}, \citenamefont {Ferrando}, \citenamefont {Johnston},\ and\
  \citenamefont {Yuan}}]{aveyard_modeling_2014}%
  \BibitemOpen
  \bibfield  {author} {\bibinfo {author} {\bibfnamefont {R.}~\bibnamefont
  {Aveyard}}, \bibinfo {author} {\bibfnamefont {R.}~\bibnamefont {Ferrando}},
  \bibinfo {author} {\bibfnamefont {R.~L.}\ \bibnamefont {Johnston}},\ and\
  \bibinfo {author} {\bibfnamefont {J.}~\bibnamefont {Yuan}},\ }\href
  {https://doi.org/10.1103/PhysRevLett.113.075501} {\bibfield  {journal}
  {\bibinfo  {journal} {Phys. Rev. Lett.}\ }\textbf {\bibinfo {volume} {113}},\
  \bibinfo {pages} {075501} (\bibinfo {year} {2014})},\ \bibinfo {note}
  {publisher: American Physical Society}\BibitemShut {NoStop}%
\bibitem [{\citenamefont {A.~Muller}\ \emph {et~al.}(2001)\citenamefont
  {A.~Muller}, \citenamefont {Edwards}, \citenamefont {J.~Kirkland},\ and\
  \citenamefont {Silcox}}]{muller_simulation_2001}%
  \BibitemOpen
  \bibfield  {author} {\bibinfo {author} {\bibfnamefont {D.}~\bibnamefont
  {A.~Muller}}, \bibinfo {author} {\bibfnamefont {B.}~\bibnamefont {Edwards}},
  \bibinfo {author} {\bibfnamefont {E.}~\bibnamefont {J.~Kirkland}},\ and\
  \bibinfo {author} {\bibfnamefont {J.}~\bibnamefont {Silcox}},\ }\href
  {https://doi.org/10.1016/S0304-3991(00)00128-5} {\bibfield  {journal}
  {\bibinfo  {journal} {Ultramicroscopy}\ }\bibinfo {series} {ISSM},\ \textbf
  {\bibinfo {volume} {86}},\ \bibinfo {pages} {371} (\bibinfo {year}
  {2001})}\BibitemShut {NoStop}%
\bibitem [{\citenamefont {Chen}\ \emph {et~al.}(2023)\citenamefont {Chen},
  \citenamefont {Kim},\ and\ \citenamefont {LeBeau}}]{chen_comparison_2023}%
  \BibitemOpen
  \bibfield  {author} {\bibinfo {author} {\bibfnamefont {X.}~\bibnamefont
  {Chen}}, \bibinfo {author} {\bibfnamefont {D.~S.}\ \bibnamefont {Kim}},\ and\
  \bibinfo {author} {\bibfnamefont {J.~M.}\ \bibnamefont {LeBeau}},\ }\href
  {https://doi.org/10.1016/j.ultramic.2022.113644} {\bibfield  {journal}
  {\bibinfo  {journal} {Ultramicroscopy}\ }\textbf {\bibinfo {volume} {244}},\
  \bibinfo {pages} {113644} (\bibinfo {year} {2023})}\BibitemShut {NoStop}%
\bibitem [{\citenamefont {Tuckerman}\ \emph {et~al.}(2006)\citenamefont
  {Tuckerman}, \citenamefont {Alejandre}, \citenamefont
  {{L{\'o}pez-Rend{\'o}n}}, \citenamefont {Jochim},\ and\ \citenamefont
  {Martyna}}]{tuckerman_liouvilleoperator_2006}%
  \BibitemOpen
  \bibfield  {author} {\bibinfo {author} {\bibfnamefont {M.~E.}\ \bibnamefont
  {Tuckerman}}, \bibinfo {author} {\bibfnamefont {J.}~\bibnamefont
  {Alejandre}}, \bibinfo {author} {\bibfnamefont {R.}~\bibnamefont
  {{L{\'o}pez-Rend{\'o}n}}}, \bibinfo {author} {\bibfnamefont {A.~L.}\
  \bibnamefont {Jochim}},\ and\ \bibinfo {author} {\bibfnamefont {G.~J.}\
  \bibnamefont {Martyna}},\ }\href
  {https://doi.org/10.1088/0305-4470/39/19/S18} {\bibfield  {journal} {\bibinfo
   {journal} {J. Phys. A: Math. Gen.}\ }\textbf {\bibinfo {volume} {39}},\
  \bibinfo {pages} {5629} (\bibinfo {year} {2006})}\BibitemShut {NoStop}%
\bibitem [{\citenamefont {Parrinello}\ and\ \citenamefont
  {Rahman}(1981)}]{parrinello_polymorphic_1981}%
  \BibitemOpen
  \bibfield  {author} {\bibinfo {author} {\bibfnamefont {M.}~\bibnamefont
  {Parrinello}}\ and\ \bibinfo {author} {\bibfnamefont {A.}~\bibnamefont
  {Rahman}},\ }\href {https://doi.org/10.1063/1.328693} {\bibfield  {journal}
  {\bibinfo  {journal} {J. Appl. Phys.}\ }\textbf {\bibinfo {volume} {52}},\
  \bibinfo {pages} {7182} (\bibinfo {year} {1981})}\BibitemShut {NoStop}%
\bibitem [{\citenamefont {Shinoda}\ \emph {et~al.}(2004)\citenamefont
  {Shinoda}, \citenamefont {Shiga},\ and\ \citenamefont
  {Mikami}}]{shinoda_rapid_2004}%
  \BibitemOpen
  \bibfield  {author} {\bibinfo {author} {\bibfnamefont {W.}~\bibnamefont
  {Shinoda}}, \bibinfo {author} {\bibfnamefont {M.}~\bibnamefont {Shiga}},\
  and\ \bibinfo {author} {\bibfnamefont {M.}~\bibnamefont {Mikami}},\ }\href
  {https://doi.org/10.1103/PhysRevB.69.134103} {\bibfield  {journal} {\bibinfo
  {journal} {Phys. Rev. B}\ }\textbf {\bibinfo {volume} {69}},\ \bibinfo
  {pages} {134103} (\bibinfo {year} {2004})}\BibitemShut {NoStop}%
\bibitem [{\citenamefont {Thompson}\ \emph {et~al.}(2015)\citenamefont
  {Thompson}, \citenamefont {Swiler}, \citenamefont {Trott}, \citenamefont
  {Foiles},\ and\ \citenamefont {Tucker}}]{thompson_spectral_2015}%
  \BibitemOpen
  \bibfield  {author} {\bibinfo {author} {\bibfnamefont {A.~P.}\ \bibnamefont
  {Thompson}}, \bibinfo {author} {\bibfnamefont {L.~P.}\ \bibnamefont
  {Swiler}}, \bibinfo {author} {\bibfnamefont {C.~R.}\ \bibnamefont {Trott}},
  \bibinfo {author} {\bibfnamefont {S.~M.}\ \bibnamefont {Foiles}},\ and\
  \bibinfo {author} {\bibfnamefont {G.~J.}\ \bibnamefont {Tucker}},\ }\href
  {https://doi.org/10.1016/j.jcp.2014.12.018} {\bibfield  {journal} {\bibinfo
  {journal} {J. Comput. Phys.}\ }\textbf {\bibinfo {volume} {285}},\ \bibinfo
  {pages} {316} (\bibinfo {year} {2015})}\BibitemShut {NoStop}%
\bibitem [{\citenamefont {Zuo}\ \emph {et~al.}(2020)\citenamefont {Zuo},
  \citenamefont {Chen}, \citenamefont {Li}, \citenamefont {Deng}, \citenamefont
  {Chen}, \citenamefont {Behler}, \citenamefont {Cs{\'a}nyi}, \citenamefont
  {Shapeev}, \citenamefont {Thompson}, \citenamefont {Wood},\ and\
  \citenamefont {Ong}}]{zuo_performance_2020}%
  \BibitemOpen
  \bibfield  {author} {\bibinfo {author} {\bibfnamefont {Y.}~\bibnamefont
  {Zuo}}, \bibinfo {author} {\bibfnamefont {C.}~\bibnamefont {Chen}}, \bibinfo
  {author} {\bibfnamefont {X.}~\bibnamefont {Li}}, \bibinfo {author}
  {\bibfnamefont {Z.}~\bibnamefont {Deng}}, \bibinfo {author} {\bibfnamefont
  {Y.}~\bibnamefont {Chen}}, \bibinfo {author} {\bibfnamefont {J.}~\bibnamefont
  {Behler}}, \bibinfo {author} {\bibfnamefont {G.}~\bibnamefont {Cs{\'a}nyi}},
  \bibinfo {author} {\bibfnamefont {A.~V.}\ \bibnamefont {Shapeev}}, \bibinfo
  {author} {\bibfnamefont {A.~P.}\ \bibnamefont {Thompson}}, \bibinfo {author}
  {\bibfnamefont {M.~A.}\ \bibnamefont {Wood}},\ and\ \bibinfo {author}
  {\bibfnamefont {S.~P.}\ \bibnamefont {Ong}},\ }\href
  {https://doi.org/10.1021/acs.jpca.9b08723} {\bibfield  {journal} {\bibinfo
  {journal} {J. Phys. Chem. A}\ }\textbf {\bibinfo {volume} {124}},\ \bibinfo
  {pages} {731} (\bibinfo {year} {2020})}\BibitemShut {NoStop}%
\bibitem [{\citenamefont {Birgeneau}\ \emph {et~al.}(1964)\citenamefont
  {Birgeneau}, \citenamefont {Cordes}, \citenamefont {Dolling},\ and\
  \citenamefont {Woods}}]{birgeneau_normal_1964}%
  \BibitemOpen
  \bibfield  {author} {\bibinfo {author} {\bibfnamefont {R.~J.}\ \bibnamefont
  {Birgeneau}}, \bibinfo {author} {\bibfnamefont {J.}~\bibnamefont {Cordes}},
  \bibinfo {author} {\bibfnamefont {G.}~\bibnamefont {Dolling}},\ and\ \bibinfo
  {author} {\bibfnamefont {A.~D.~B.}\ \bibnamefont {Woods}},\ }\href
  {https://doi.org/10.1103/PhysRev.136.A1359} {\bibfield  {journal} {\bibinfo
  {journal} {Phys. Rev.}\ }\textbf {\bibinfo {volume} {136}},\ \bibinfo {pages}
  {A1359} (\bibinfo {year} {1964})}\BibitemShut {NoStop}%
\bibitem [{\citenamefont {Lee}\ \emph {et~al.}(1993)\citenamefont {Lee},
  \citenamefont {Vanderbilt}, \citenamefont {Laasonen}, \citenamefont {Car},\
  and\ \citenamefont {Parrinello}}]{lee_initio_1993}%
  \BibitemOpen
  \bibfield  {author} {\bibinfo {author} {\bibfnamefont {C.}~\bibnamefont
  {Lee}}, \bibinfo {author} {\bibfnamefont {D.}~\bibnamefont {Vanderbilt}},
  \bibinfo {author} {\bibfnamefont {K.}~\bibnamefont {Laasonen}}, \bibinfo
  {author} {\bibfnamefont {R.}~\bibnamefont {Car}},\ and\ \bibinfo {author}
  {\bibfnamefont {M.}~\bibnamefont {Parrinello}},\ }\href
  {https://doi.org/10.1103/PhysRevB.47.4863} {\bibfield  {journal} {\bibinfo
  {journal} {Phys. Rev. B}\ }\textbf {\bibinfo {volume} {47}},\ \bibinfo
  {pages} {4863} (\bibinfo {year} {1993})}\BibitemShut {NoStop}%
\bibitem [{\citenamefont {Carreras}\ \emph {et~al.}(2017)\citenamefont
  {Carreras}, \citenamefont {Togo},\ and\ \citenamefont
  {Tanaka}}]{carreras_dynaphopy_2017}%
  \BibitemOpen
  \bibfield  {author} {\bibinfo {author} {\bibfnamefont {A.}~\bibnamefont
  {Carreras}}, \bibinfo {author} {\bibfnamefont {A.}~\bibnamefont {Togo}},\
  and\ \bibinfo {author} {\bibfnamefont {I.}~\bibnamefont {Tanaka}},\ }\href
  {https://doi.org/10.1016/j.cpc.2017.08.017} {\bibfield  {journal} {\bibinfo
  {journal} {Comput. Phys. Commun.}\ }\textbf {\bibinfo {volume} {221}},\
  \bibinfo {pages} {221} (\bibinfo {year} {2017})}\BibitemShut {NoStop}%
\bibitem [{\citenamefont {Qi}\ \emph {et~al.}(2020)\citenamefont {Qi},
  \citenamefont {Ma}, \citenamefont {Zhao}, \citenamefont {Cheng},
  \citenamefont {Jiang}, \citenamefont {Lu}, \citenamefont {Jiang},
  \citenamefont {Qian}, \citenamefont {Wang}, \citenamefont {Zhang},
  \citenamefont {Zhu}, \citenamefont {Zou}, \citenamefont {Wan}, \citenamefont
  {Xiang},\ and\ \citenamefont {Zhang}}]{qi_breaking_2020}%
  \BibitemOpen
  \bibfield  {author} {\bibinfo {author} {\bibfnamefont {F.}~\bibnamefont
  {Qi}}, \bibinfo {author} {\bibfnamefont {Z.}~\bibnamefont {Ma}}, \bibinfo
  {author} {\bibfnamefont {L.}~\bibnamefont {Zhao}}, \bibinfo {author}
  {\bibfnamefont {Y.}~\bibnamefont {Cheng}}, \bibinfo {author} {\bibfnamefont
  {W.}~\bibnamefont {Jiang}}, \bibinfo {author} {\bibfnamefont
  {C.}~\bibnamefont {Lu}}, \bibinfo {author} {\bibfnamefont {T.}~\bibnamefont
  {Jiang}}, \bibinfo {author} {\bibfnamefont {D.}~\bibnamefont {Qian}},
  \bibinfo {author} {\bibfnamefont {Z.}~\bibnamefont {Wang}}, \bibinfo {author}
  {\bibfnamefont {W.}~\bibnamefont {Zhang}}, \bibinfo {author} {\bibfnamefont
  {P.}~\bibnamefont {Zhu}}, \bibinfo {author} {\bibfnamefont {X.}~\bibnamefont
  {Zou}}, \bibinfo {author} {\bibfnamefont {W.}~\bibnamefont {Wan}}, \bibinfo
  {author} {\bibfnamefont {D.}~\bibnamefont {Xiang}},\ and\ \bibinfo {author}
  {\bibfnamefont {J.}~\bibnamefont {Zhang}},\ }\href
  {https://doi.org/10.1103/PhysRevLett.124.134803} {\bibfield  {journal}
  {\bibinfo  {journal} {Phys. Rev. Lett.}\ }\textbf {\bibinfo {volume} {124}},\
  \bibinfo {pages} {134803} (\bibinfo {year} {2020})}\BibitemShut {NoStop}%
\bibitem [{\citenamefont {{Sokolowski-Tinten}}\ \emph
  {et~al.}(2003)\citenamefont {{Sokolowski-Tinten}}, \citenamefont {Blome},
  \citenamefont {Blums}, \citenamefont {Cavalleri}, \citenamefont {Dietrich},
  \citenamefont {Tarasevitch}, \citenamefont {Uschmann}, \citenamefont
  {F{\"o}rster}, \citenamefont {Kammler}, \citenamefont {{Horn-von-Hoegen}},\
  and\ \citenamefont {{von der Linde}}}]{sokolowski-tinten_femtosecond_2003}%
  \BibitemOpen
  \bibfield  {author} {\bibinfo {author} {\bibfnamefont {K.}~\bibnamefont
  {{Sokolowski-Tinten}}}, \bibinfo {author} {\bibfnamefont {C.}~\bibnamefont
  {Blome}}, \bibinfo {author} {\bibfnamefont {J.}~\bibnamefont {Blums}},
  \bibinfo {author} {\bibfnamefont {A.}~\bibnamefont {Cavalleri}}, \bibinfo
  {author} {\bibfnamefont {C.}~\bibnamefont {Dietrich}}, \bibinfo {author}
  {\bibfnamefont {A.}~\bibnamefont {Tarasevitch}}, \bibinfo {author}
  {\bibfnamefont {I.}~\bibnamefont {Uschmann}}, \bibinfo {author}
  {\bibfnamefont {E.}~\bibnamefont {F{\"o}rster}}, \bibinfo {author}
  {\bibfnamefont {M.}~\bibnamefont {Kammler}}, \bibinfo {author} {\bibfnamefont
  {M.}~\bibnamefont {{Horn-von-Hoegen}}},\ and\ \bibinfo {author}
  {\bibfnamefont {D.}~\bibnamefont {{von der Linde}}},\ }\href
  {https://doi.org/10.1038/nature01490} {\bibfield  {journal} {\bibinfo
  {journal} {Nature}\ }\textbf {\bibinfo {volume} {422}},\ \bibinfo {pages}
  {287} (\bibinfo {year} {2003})}\BibitemShut {NoStop}%
\bibitem [{\citenamefont {Harmand}\ \emph {et~al.}(2013)\citenamefont
  {Harmand}, \citenamefont {Coffee}, \citenamefont {Bionta}, \citenamefont
  {Chollet}, \citenamefont {French}, \citenamefont {Zhu}, \citenamefont
  {Fritz}, \citenamefont {Lemke}, \citenamefont {Medvedev}, \citenamefont
  {Ziaja}, \citenamefont {Toleikis},\ and\ \citenamefont
  {Cammarata}}]{harmand_achieving_2013}%
  \BibitemOpen
  \bibfield  {author} {\bibinfo {author} {\bibfnamefont {M.}~\bibnamefont
  {Harmand}}, \bibinfo {author} {\bibfnamefont {R.}~\bibnamefont {Coffee}},
  \bibinfo {author} {\bibfnamefont {M.~R.}\ \bibnamefont {Bionta}}, \bibinfo
  {author} {\bibfnamefont {M.}~\bibnamefont {Chollet}}, \bibinfo {author}
  {\bibfnamefont {D.}~\bibnamefont {French}}, \bibinfo {author} {\bibfnamefont
  {D.}~\bibnamefont {Zhu}}, \bibinfo {author} {\bibfnamefont {D.~M.}\
  \bibnamefont {Fritz}}, \bibinfo {author} {\bibfnamefont {H.~T.}\ \bibnamefont
  {Lemke}}, \bibinfo {author} {\bibfnamefont {N.}~\bibnamefont {Medvedev}},
  \bibinfo {author} {\bibfnamefont {B.}~\bibnamefont {Ziaja}}, \bibinfo
  {author} {\bibfnamefont {S.}~\bibnamefont {Toleikis}},\ and\ \bibinfo
  {author} {\bibfnamefont {M.}~\bibnamefont {Cammarata}},\ }\href
  {https://doi.org/10.1038/nphoton.2013.11} {\bibfield  {journal} {\bibinfo
  {journal} {Nat. Photon.}\ }\textbf {\bibinfo {volume} {7}},\ \bibinfo {pages}
  {215} (\bibinfo {year} {2013})}\BibitemShut {NoStop}%
\bibitem [{\citenamefont {Fritz}\ \emph {et~al.}(2007)\citenamefont {Fritz},
  \citenamefont {Reis}, \citenamefont {Adams}, \citenamefont {Akre},
  \citenamefont {Arthur}, \citenamefont {Blome}, \citenamefont {Bucksbaum},
  \citenamefont {Cavalieri}, \citenamefont {Engemann}, \citenamefont {Fahy},
  \citenamefont {Falcone}, \citenamefont {Fuoss}, \citenamefont {Gaffney},
  \citenamefont {George}, \citenamefont {Hajdu}, \citenamefont {Hertlein},
  \citenamefont {Hillyard}, \citenamefont {{Horn-von Hoegen}}, \citenamefont
  {Kammler}, \citenamefont {Kaspar}, \citenamefont {Kienberger}, \citenamefont
  {Krejcik}, \citenamefont {Lee}, \citenamefont {Lindenberg}, \citenamefont
  {McFarland}, \citenamefont {Meyer}, \citenamefont {Montagne}, \citenamefont
  {Murray}, \citenamefont {Nelson}, \citenamefont {Nicoul}, \citenamefont
  {Pahl}, \citenamefont {Rudati}, \citenamefont {Schlarb}, \citenamefont
  {Siddons}, \citenamefont {{Sokolowski-Tinten}}, \citenamefont {Tschentscher},
  \citenamefont {{von der Linde}},\ and\ \citenamefont
  {Hastings}}]{fritz_ultrafast_2007}%
  \BibitemOpen
  \bibfield  {author} {\bibinfo {author} {\bibfnamefont {D.~M.}\ \bibnamefont
  {Fritz}}, \bibinfo {author} {\bibfnamefont {D.~A.}\ \bibnamefont {Reis}},
  \bibinfo {author} {\bibfnamefont {B.}~\bibnamefont {Adams}}, \bibinfo
  {author} {\bibfnamefont {R.~A.}\ \bibnamefont {Akre}}, \bibinfo {author}
  {\bibfnamefont {J.}~\bibnamefont {Arthur}}, \bibinfo {author} {\bibfnamefont
  {C.}~\bibnamefont {Blome}}, \bibinfo {author} {\bibfnamefont {P.~H.}\
  \bibnamefont {Bucksbaum}}, \bibinfo {author} {\bibfnamefont {A.~L.}\
  \bibnamefont {Cavalieri}}, \bibinfo {author} {\bibfnamefont {S.}~\bibnamefont
  {Engemann}}, \bibinfo {author} {\bibfnamefont {S.}~\bibnamefont {Fahy}},
  \bibinfo {author} {\bibfnamefont {R.~W.}\ \bibnamefont {Falcone}}, \bibinfo
  {author} {\bibfnamefont {P.~H.}\ \bibnamefont {Fuoss}}, \bibinfo {author}
  {\bibfnamefont {K.~J.}\ \bibnamefont {Gaffney}}, \bibinfo {author}
  {\bibfnamefont {M.~J.}\ \bibnamefont {George}}, \bibinfo {author}
  {\bibfnamefont {J.}~\bibnamefont {Hajdu}}, \bibinfo {author} {\bibfnamefont
  {M.~P.}\ \bibnamefont {Hertlein}}, \bibinfo {author} {\bibfnamefont {P.~B.}\
  \bibnamefont {Hillyard}}, \bibinfo {author} {\bibfnamefont {M.}~\bibnamefont
  {{Horn-von Hoegen}}}, \bibinfo {author} {\bibfnamefont {M.}~\bibnamefont
  {Kammler}}, \bibinfo {author} {\bibfnamefont {J.}~\bibnamefont {Kaspar}},
  \bibinfo {author} {\bibfnamefont {R.}~\bibnamefont {Kienberger}}, \bibinfo
  {author} {\bibfnamefont {P.}~\bibnamefont {Krejcik}}, \bibinfo {author}
  {\bibfnamefont {S.~H.}\ \bibnamefont {Lee}}, \bibinfo {author} {\bibfnamefont
  {A.~M.}\ \bibnamefont {Lindenberg}}, \bibinfo {author} {\bibfnamefont
  {B.}~\bibnamefont {McFarland}}, \bibinfo {author} {\bibfnamefont
  {D.}~\bibnamefont {Meyer}}, \bibinfo {author} {\bibfnamefont
  {T.}~\bibnamefont {Montagne}}, \bibinfo {author} {\bibfnamefont {E.~D.}\
  \bibnamefont {Murray}}, \bibinfo {author} {\bibfnamefont {A.~J.}\
  \bibnamefont {Nelson}}, \bibinfo {author} {\bibfnamefont {M.}~\bibnamefont
  {Nicoul}}, \bibinfo {author} {\bibfnamefont {R.}~\bibnamefont {Pahl}},
  \bibinfo {author} {\bibfnamefont {J.}~\bibnamefont {Rudati}}, \bibinfo
  {author} {\bibfnamefont {H.}~\bibnamefont {Schlarb}}, \bibinfo {author}
  {\bibfnamefont {D.~P.}\ \bibnamefont {Siddons}}, \bibinfo {author}
  {\bibfnamefont {K.}~\bibnamefont {{Sokolowski-Tinten}}}, \bibinfo {author}
  {\bibfnamefont {{\relax Th}.}~\bibnamefont {Tschentscher}}, \bibinfo {author}
  {\bibfnamefont {D.}~\bibnamefont {{von der Linde}}},\ and\ \bibinfo {author}
  {\bibfnamefont {J.~B.}\ \bibnamefont {Hastings}},\ }\href
  {https://doi.org/10.1126/science.1135009} {\bibfield  {journal} {\bibinfo
  {journal} {Science}\ }\textbf {\bibinfo {volume} {315}},\ \bibinfo {pages}
  {633} (\bibinfo {year} {2007})}\BibitemShut {NoStop}%
\bibitem [{\citenamefont {Zhao}\ \emph {et~al.}(2021)\citenamefont {Zhao},
  \citenamefont {Wu}, \citenamefont {Wang}, \citenamefont {Tang}, \citenamefont
  {Zou}, \citenamefont {Jiang}, \citenamefont {Zhu}, \citenamefont {Xiang},\
  and\ \citenamefont {Zhang}}]{zhao_noninvasive_2021}%
  \BibitemOpen
  \bibfield  {author} {\bibinfo {author} {\bibfnamefont {L.}~\bibnamefont
  {Zhao}}, \bibinfo {author} {\bibfnamefont {J.}~\bibnamefont {Wu}}, \bibinfo
  {author} {\bibfnamefont {Z.}~\bibnamefont {Wang}}, \bibinfo {author}
  {\bibfnamefont {H.}~\bibnamefont {Tang}}, \bibinfo {author} {\bibfnamefont
  {X.}~\bibnamefont {Zou}}, \bibinfo {author} {\bibfnamefont {T.}~\bibnamefont
  {Jiang}}, \bibinfo {author} {\bibfnamefont {P.}~\bibnamefont {Zhu}}, \bibinfo
  {author} {\bibfnamefont {D.}~\bibnamefont {Xiang}},\ and\ \bibinfo {author}
  {\bibfnamefont {J.}~\bibnamefont {Zhang}},\ }\href
  {https://doi.org/10.1063/4.0000113} {\bibfield  {journal} {\bibinfo
  {journal} {Struct. Dyn.}\ }\textbf {\bibinfo {volume} {8}},\ \bibinfo {pages}
  {044303} (\bibinfo {year} {2021})}\BibitemShut {NoStop}%
\bibitem [{\citenamefont {Hellsvik}\ \emph {et~al.}(2019)\citenamefont
  {Hellsvik}, \citenamefont {Thonig}, \citenamefont {Modin}, \citenamefont
  {Iu\ifmmode~\mbox{\c{s}}\else \c{s}\fi{}an}, \citenamefont {Bergman},
  \citenamefont {Eriksson}, \citenamefont {Bergqvist},\ and\ \citenamefont
  {Delin}}]{hellsvik_general_2019}%
  \BibitemOpen
  \bibfield  {author} {\bibinfo {author} {\bibfnamefont {J.}~\bibnamefont
  {Hellsvik}}, \bibinfo {author} {\bibfnamefont {D.}~\bibnamefont {Thonig}},
  \bibinfo {author} {\bibfnamefont {K.}~\bibnamefont {Modin}}, \bibinfo
  {author} {\bibfnamefont {D.}~\bibnamefont {Iu\ifmmode~\mbox{\c{s}}\else
  \c{s}\fi{}an}}, \bibinfo {author} {\bibfnamefont {A.}~\bibnamefont
  {Bergman}}, \bibinfo {author} {\bibfnamefont {O.}~\bibnamefont {Eriksson}},
  \bibinfo {author} {\bibfnamefont {L.}~\bibnamefont {Bergqvist}},\ and\
  \bibinfo {author} {\bibfnamefont {A.}~\bibnamefont {Delin}},\ }\href
  {https://doi.org/10.1103/PhysRevB.99.104302} {\bibfield  {journal} {\bibinfo
  {journal} {Phys. Rev. B}\ }\textbf {\bibinfo {volume} {99}},\ \bibinfo
  {pages} {104302} (\bibinfo {year} {2019})}\BibitemShut {NoStop}%
\bibitem [{\citenamefont {Edstr\"om}\ \emph {et~al.}(2016)\citenamefont
  {Edstr\"om}, \citenamefont {Lubk},\ and\ \citenamefont
  {Rusz}}]{edstrom_elastic_2016}%
  \BibitemOpen
  \bibfield  {author} {\bibinfo {author} {\bibfnamefont {A.}~\bibnamefont
  {Edstr\"om}}, \bibinfo {author} {\bibfnamefont {A.}~\bibnamefont {Lubk}},\
  and\ \bibinfo {author} {\bibfnamefont {J.}~\bibnamefont {Rusz}},\ }\href
  {https://doi.org/10.1103/PhysRevLett.116.127203} {\bibfield  {journal}
  {\bibinfo  {journal} {Phys. Rev. Lett.}\ }\textbf {\bibinfo {volume} {116}},\
  \bibinfo {pages} {127203} (\bibinfo {year} {2016})}\BibitemShut {NoStop}%
\bibitem [{\citenamefont {Castellanos-Reyes}\ \emph {et~al.}(2023)\citenamefont
  {Castellanos-Reyes}, \citenamefont {Zeiger}, \citenamefont {Bergman},
  \citenamefont {Kepaptsoglou}, \citenamefont {Ramasse}, \citenamefont
  {Idrobo},\ and\ \citenamefont {Rusz}}]{Castellanos-Reyes_unveiling_2023}%
  \BibitemOpen
  \bibfield  {author} {\bibinfo {author} {\bibfnamefont {J.~{\'{A}}.}\
  \bibnamefont {Castellanos-Reyes}}, \bibinfo {author} {\bibfnamefont
  {P.}~\bibnamefont {Zeiger}}, \bibinfo {author} {\bibfnamefont
  {A.}~\bibnamefont {Bergman}}, \bibinfo {author} {\bibfnamefont
  {D.}~\bibnamefont {Kepaptsoglou}}, \bibinfo {author} {\bibfnamefont {Q.~M.}\
  \bibnamefont {Ramasse}}, \bibinfo {author} {\bibfnamefont {J.~C.}\
  \bibnamefont {Idrobo}},\ and\ \bibinfo {author} {\bibfnamefont
  {J.}~\bibnamefont {Rusz}},\ }\href
  {https://doi.org/10.1103/PhysRevB.108.134435} {\bibfield  {journal} {\bibinfo
   {journal} {Phys. Rev. B}\ }\textbf {\bibinfo {volume} {108}},\ \bibinfo
  {pages} {134435} (\bibinfo {year} {2023})}\BibitemShut {NoStop}%
\bibitem [{\citenamefont {Bloomfield}(2004)}]{bloomfield_fourier_2004}%
  \BibitemOpen
  \bibfield  {author} {\bibinfo {author} {\bibfnamefont {P.}~\bibnamefont
  {Bloomfield}},\ }\href@noop {} {\emph {\bibinfo {title} {Fourier {Analysis}
  of {Time} {Series}: {An} {Introduction}}}}\ (\bibinfo  {publisher} {John
  Wiley \& Sons},\ \bibinfo {year} {2004})\BibitemShut {NoStop}%
\end{thebibliography}%

\end{document}